\documentclass[3p]{elsarticle}
\usepackage{amsmath,amsfonts,amssymb,mathrsfs,stmaryrd,pifont,bbm,mathtools,multirow,color,amsthm}
\usepackage{lineno,hyperref,algorithmicx}
\usepackage[ruled]{algorithm2e}
\usepackage{makecell}
\usepackage{multirow,threeparttable}
\usepackage{amsmath,amsfonts}
\usepackage{framed} 
\usepackage{multicol} 
\usepackage{nomencl} 
\usepackage{cuted}
\usepackage{acro}
\usepackage{longtable}
\modulolinenumbers[5]
\journal{Renewable and Sustainable Energy Reviews on 09-Jan-2024}

\allowdisplaybreaks[4]

\DeclareAcronym{LIBs}{
  short = LIBs,
  long  = Lithium-ion batteries
}

\DeclareAcronym{BMS}{
  short = BMS,
  long  = Battery management system
}

\DeclareAcronym{EV}{
  short = EV,
  long  = Electric vehicle
}

\DeclareAcronym{SC}{
  short = SC,
  long  = Short circuit
}

\DeclareAcronym{ISC}{
  short = ISC,
  long  = Internal short circuit
}

\DeclareAcronym{ESC}{
  short = ESC,
  long  = External short circuit
}

\DeclareAcronym{EM}{
  short = EM,
  long  = Electrochemical model
}

\DeclareAcronym{ECM}{
  short = ECM,
  long  = Electrical equivalent circuit model
}

\DeclareAcronym{PDE}{
  short = PDE,
  long  = Partial differential equation
}

\DeclareAcronym{SP}{
  short = SP,
  long  = Single particle
}

\DeclareAcronym{EIS}{
  short = EIS,
  long  = Electrochemical impedance spectroscopy
}

\DeclareAcronym{IOM}{
  short = IOM,
  long  = Integral order model
}

\DeclareAcronym{FOM}{
  short = FOM,
  long  = Fractional order model
}

\DeclareAcronym{OCV}{
  short = OCV,
  long  = Open circuit voltage
}

\DeclareAcronym{KF}{
  short = KF,
  long  = Kalman filter
}

\DeclareAcronym{RLS}{
  short = RLS,
  long  = Recursive least square
}

\DeclareAcronym{GA}{
  short = GA,
  long  = Genetic algorithm
}

\DeclareAcronym{PSO}{
  short = PSO,
  long  = Particle swarm optimization
}

\DeclareAcronym{SVO}{
  short = SVO,
  long  = Set-valued observer
}

\DeclareAcronym{LO}{
  short = LO,
  long  = Luenberger observer
}

\DeclareAcronym{SMO}{
  short = SMO,
  long  = Sliding mode observer
}

\DeclareAcronym{P2D}{
  short = P2D,
  long  = Pseudo-two-dimensional
}

\DeclareAcronym{PIO}{
  short = PIO,
  long  = Proportional integral observer
}

\DeclareAcronym{HIO}{
  short = HIO,
  long  = $H_\infty$ observer
}

\DeclareAcronym{SOH}{
  short = SOH,
  long  = State of health
}

\DeclareAcronym{SOC}{
  short = SOC,
  long  = State of charge
}

\DeclareAcronym{CPE}{
  short = CPE,
  long  = Constant phase element
}

\DeclareAcronym{RCC}{
  short = RCC,
  long  = Remaining charge capacity
}

\begin{document}

\begin{frontmatter}
\title{Recent Advances in Model-Based Fault Diagnosis for Lithium-Ion Batteries: A Comprehensive Review}

\author{Yiming Xu}
\ead{103092673@student.swin.edu.au}

\author{Xiaohua Ge}
\ead{xge@swin.com}

\author{Ruohan Guo}
\ead{rguo@swin.com}

\author{Weixiang Shen\corref{cor1}}
\ead{wshen@swin.com}
\cortext[cor1]{Corresponding author}

\address{School of Science, Computing and Engineering Technologies, Swinburne University of Technology, Hawthorn, 3122, Victoria, Australia}

\begin{abstract}
Lithium-ion batteries (LIBs) have found wide applications in a variety of fields such as electrified transportation, stationary storage and portable electronics devices. A battery management system (BMS) is critical to ensure the reliability, efficiency and longevity of LIBs. Recent research has witnessed the emergence of model-based fault diagnosis methods in advanced BMSs. This paper provides a comprehensive review on the model-based fault diagnosis methods for LIBs. First, the widely explored battery models in the existing literature are classified into physics-based electrochemical models and electrical equivalent circuit models. Second, a general state-space representation that describes electrical dynamics of a faulty battery is presented.The formulation of the state vectors and the identification of the parameter matrices are then elaborated.
Third, the fault mechanisms of both battery faults (incl. overcharege/overdischarge faults, connection faults, short circuit faults) and sensor faults (incl. voltage sensor faults and current sensor faults) are discussed. Furthermore, different types of modeling uncertainties, such as modeling errors and measurement noises, aging effects, measurement outliers, are elaborated. An emphasis is then placed on the observer design (incl. online state observers and offline state observers). The algorithm implementation of typical state observers for battery fault diagnosis is also put forward. Finally, discussion and outlook are offered to envision some possible future research directions.
\end{abstract}

\begin{keyword}
Lithium-ion battery, Battery management, Battery modeling, State estimation, Fault diagnosis.
\end{keyword}

\end{frontmatter}

\begin{longtable}{ll}
\textbf{Acronym} & \textbf{Full Form} \\
  \endhead
  \ac{LIBs} & Lithium-ion batteries \\
  \ac{EV} & Electric vehicle \\
  \ac{BMS} & Battery management system \\
  \ac{EM} & Electrochemical model \\
  \ac{ECM} & Electrical equivalent circuit model \\
  \ac{SC} & Short circuit \\
  \ac{ISC} & Internal short circuit\\
  \ac{ESC} & External short circuit\\
  \ac{PDE} & Partial differential equation \\
  \ac{P2D} & Pseudo-two-dimensional\\
  \ac{SP} & Single particle \\
  \ac{EIS} & Electrochemical impedance spectroscopy \\
  \ac{IOM} & Integral order model \\
  \ac{FOM} & Fractional order model \\
  \ac{SOC} & State of charge\\
  \ac{OCV} & Open circuit voltage \\
  \ac{CPE} & Constant phase element\\
  \ac{SOH} & State of health \\
  \ac{KF} & Kalman filter \\
  \ac{RLS} & Recursive least square \\
  \ac{GA} & Genetic algorithm \\
  \ac{PSO} & Particle swarm optimization \\
  \ac{SVO} & Set-valued observer \\
  \ac{LO} & Luenberger observer \\
  \ac{SMO} & Sliding mode observer \\
  \ac{PIO} & Proportional integral observer \\
  \ac{HIO} & $H_\infty$ observer \\
  \ac{RCC} & Remaining charge capacity
\end{longtable}
\section{Introduction}
Lithium-ion batteries (LIBs) have been emerging as one of the most promising energy storage systems in electric vehicles (EVs), renewable energy systems and portable electronic devices due to their high energy density and long life span. However, potential risks coming from abusive operations and harsh environments pose threats to the safety of LIBs \cite{JLJES2022}. To ensure the normal operation and expected service life of batteries, it is essential to implement effective battery management strategies \cite{MSJCP2021, HDRe2021}. The battery management system (BMS) encompasses a range of functions, including condition monitoring, thermal management, cell balancing, state estimation and fault diagnosis \cite{RGVehicles2022, JTESM2023}. Among these, fault diagnosis plays a pivotal role in preserving the health and reliability of battery systems \cite{RXIscience2020} as even a minor fault could eventually lead severe damage to LIBs \cite{WLEnerg2022, MaAEnergy2021}. Hence, developing advanced and intelligent fault diagnosis algorithms for early detection of battery faults has become a hot research topic.

Owing to the narrow operational temperature range and the unpredictable working environment, LIBs may be susceptible to thermal runaway when they are exposed to conditions such as overheat, overcharging, overdischarging or mechanical damage.
Noteworthy incidents, such as an electric bus bursting into flames after colliding with a guardrail on a highway in 2020 \cite{ANEW}, and a Tesla Model S catching fire after a high-speed impact to a wall and tree in 2013 \cite{TSHS}, underscore the vulnerability of LIBs to hazardous events.
In addition to the possible damage to battery modules caused by collisions, the combustion of EVs during the charging or operating process accounts for a large part of electric vehicle fire accidents. For instance, a Tesla Model S spontaneously ignited during recharging \cite{TeslaS} and a parked electric MG caught fire due to a damaged EV battery \cite{MGDrive}.
These incidents highlight the critical significance of fault diagnosis strategies to mitigate the risks assocated with LIBs in various operaing conditions.

Up till now, several reviews on fault diagnosis of LIBs have been available in the literature \cite{ WYRenew2020, XLESM2021, GZRenew2021, QYCWET,YYAAE2023}.
The differences in terms of literature covering period and technical focus between this study and some existing reviews on LIB fault diagnosis are briefly outlined in Table \ref{tab1}. It should be mentioned that the existing reviews on LIB fault diagnosis methods can be generally categorized into two types: model-based methods and data-driven methods, where the former commonly regard the abnormity detection as classification tasks through either supervised or unsupervised \cite{KhaoulaARC2016} while the latter formalize the battery physical understanding into analytical and mathmatical models based on a priori knowledge \cite{XLAutomotive2023}, but with discernible variations in focal points. In this review, we focus on model-based methods for LIB fault detection, fault identification and fault estimation. In particular, we offer 1) a thorough elucidation of a general state-space representation for a faulty battery model, involving the detailed formulation of the battery system state vector and the identification of system parameters; 2) an elaborate exposition of design principles underlying various model-based state observers and their implementation algorithms; and 3) a detailed analysis of the complete fault diagnosis procedure, including fault detection, fault identification
and fault estimation. 

\begin{table}[!t]
\begin{center}
\begin{threeparttable}[b]
\caption{Differences between different reviews on lithium-ion battery fault diagnosis}
\label{tab1}
\begin{tabular}{ l  l  l}
\hline
\hline
References & Literature covered until\tnote{1} & Focus\\
\hline
\cite{WYRenew2020} & Oct. 2020 & {\makecell[l]{Estimation approaches of state of charge, state of power,\\ state of energy}}\\
\hline
\cite{XLESM2021} & Mar. 2021 & {\makecell[l]{Mechanism, modeling and diagnosis of \\internal short circuit (ISC)}}\\
\hline
\cite{GZRenew2021} &  May 2021 & Performance analysis of ISC experimental approaches \\
\hline
\cite{QYCWET} & Jul. 2023 & {\makecell[l]{Battery measurement data characteristics\\ in different application scenarios}}\\
\hline
\cite{YYAAE2023} & Sep. 2023 & {\makecell[l]{Failure mechanism and evolution of \\ long-medium-short graded faults}}\\
\hline
This review & Mar. 2024 & {\makecell[l]{Model-based observer design and algorithm implementation,\\  model-based fault diagnosis}}\\
\hline
\hline
\end{tabular}
\begin{tablenotes}
     \item[1] Based on the acceptance month of each published review and the submission month of this review.
   \end{tablenotes}
  \end{threeparttable}
\end{center}
\end{table}

The rest of this review is organized as follows. Section 2 introduces the battery models including physics-based electrochemical models (EMs) and electrical equivalent circuit models (ECMs). Section 3 presents a general state-space representation for the ECM, where the formulation of state vectors and derivation of parameter matrices are also provided. In Section 4, the fault mechanisms are outlined. Section 5 elaborates the sources of uncertainties and discusses their impacts on state estimation. Section 6 surveys the existing state observers, online and offline, for state estimation and prediction purposes, and also entails some typical implementation algorithms. Section 7 reviews the comprehensive fault diagnosis procedure, which includes fault detection, fault identification and fault estimation. Some potential future directions and conclusion are provided in Section 8 and Section 9, respectively.

\section{Battery Modeling}

Given the intricate multi-layer internal structure of a LIB and the electrothermal coupling effect caused by faults,
establishing a well-balanced battery model between fidelity and complexity poses a critical challenge to battery fault diagnosis. The battery models employed in the literature mainly fall into the following two categories: EMs and ECMs.

\subsection{Physics-based Electrochemical Models}

EMs use partial differential equations (PDEs) to depict battery behavior, taking into account factors such as electrolyte concentration, electrode size, and internal electrochemical processes \cite{BLXTElectro2020}.
Newman and Doyle et al. \cite{MDJelectrochem1993} presented the pseudo-two-dimensional (P2D) model based on the porous electrode theory and the concentration solution theory. The model describes mass and charge conservation in solids and electrolytes via strongly coupled PDEs and describes the electrochemical kinetics via the Butler-Volmer equation. Although EMs can accurately predict battery dynamics, they entail high computation burden and time-consuming parameter identification.
For simplification, mathematical order reduction methods, including the orthogonal decomposition \cite{CLJES2010, NPWCJES2001}, the polynomial approximation \cite{SVJES2001}, the Galerkin's method \cite{DTSJPS2012, FGJES2016} and the Pade approximation \cite{JECJES2011}, have been proposed to improve the computation efficiency of the P2D model \cite{XLEnergy2016}.
In addition, physical simplification methods directly targeting the model structure have been exploited in the literature, among which the single particle (SP) model has been deemed as the most representative one \cite{DZJES2000, CLEnergy2021}.
In particular, the SP model assumes each electrode as one spherical particle and neglects Li concentration and potential gradients within electrolyte phase. On the basis of the SP, the extended SP model \cite{XHJPS2015, XHMOJPS2015, ZDE2018, TREnergy2105, LRGZJES2021}
taking into account electrolyte dynamics to improve the prediction under high C-rates, and the multi-particle model \cite{MSJES2011, MFDCJES2011} including the nonuniform charge transfer reaction inside electrodes are further introduced.

\subsection{Electrical Equivalent Circuit Models}

Due to the simplified model structure and easy parameter identification, ECMs have been widely investigated for online fault diagnosis. In ECMs, basic electrical components such as voltage sources, resistors, and capacitors are adopted to imitate battery behavior. According to the difference in fitting electrochemical impedance spectroscopy (EIS), the existing ECMs can be further divided into two groups: integral-order models (IOMs) and fractional-order models (FOMs).

\subsubsection{Integral-Order Models}

The simplest IOM is the Rint model which is composed of a state of charge (SOC)-dependent open circuit voltage (OCV) source ($U_{oc}$) in series with a constant internal resistance ($R_{0}$). For example, 
Shen et al. \cite{PSTVT2018} presented the Rint model to predict charge/discharge power and describe electrical behavior with good simplicity.
However, it fails to exhibit internal polarization and diffusion phenomena, which  results in poor SOC estimation. Based on the Rint model, pairs of parallel resistor-capacitors (RC) are connected into the model to better reflect dynamic electrical characteristics. Ding et al. \cite{XDAE2019} proposed an improved Thevenin model taking into account temperature influence on the OCV of a battery. As validated through experimental results, the model can demonstrate its capability to emulate the terminal voltage variation with high fidelity.
Cambron et al. \cite{DCTVT2017} employed a second-order RC circuit for battery modeling with one representing fast dynamics of relaxation effects and the other representing slow dynamic of relaxation effect. The results showed that the maximum modeling error in terminal voltage is less than 20mV.
According to the circuit theory, the continuous-time domain model equation and parameters of three commonly applied IOMs are specified in Table \ref{tab2}.

\begin{table*}[!htb]
\begin{center}
\caption{Representative continuous-time integral-order ECMs and their circuit diagrams}
\label{tab2}
\begin{tabular}{ l  l  l  l}
\hline
\hline
Circuit diagrams & References & Model equations & Model parameters\\
\hline
{\makecell[l]{Rint model \\
\begin{minipage}[b]{0.2\columnwidth}
		\centering
		\raisebox{-.5\height}{\includegraphics[width=\linewidth]{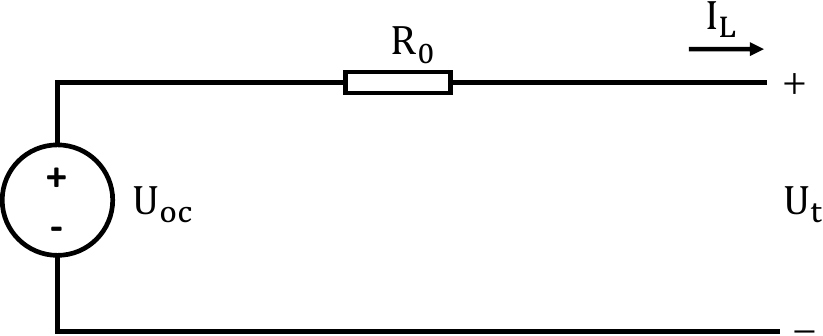}}
\end{minipage}}}
& {\makecell[l]{\cite{MSEnergies2018, XDJES2022}\\ \cite{JXEnergy2020}}}& $ U_t(t) = U_{oc}(t, SOC) - I(t)R_0(t)$&  $U_{oc}(t, SOC)$, $R_0(t)$\\
{\makecell[l]{Thevenin model \\
\begin{minipage}[b]{0.2\columnwidth}
		\centering
		\raisebox{-.5\height}{\includegraphics[width=\linewidth]{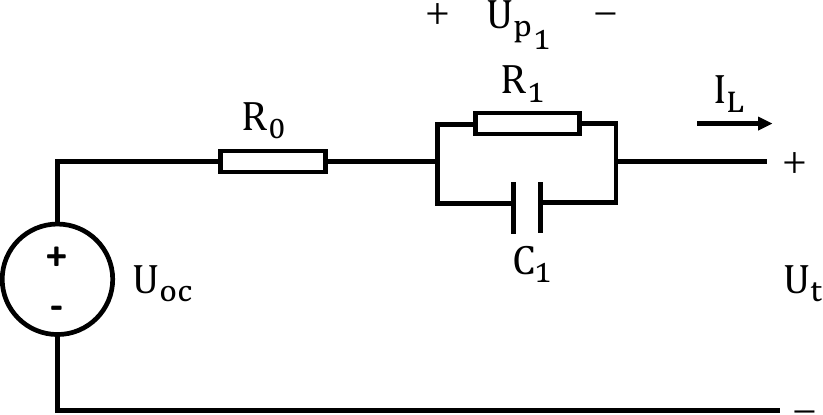}}
\end{minipage}}} & {\makecell[l]{\cite{JLJES2023,DLEnergy2023}\\ \cite{DLEnergy2023, ZSAE2022}}} & {\makecell[l]{$U_t(t) = U_{oc}(t, SOC)$ \\$- I(t)R_0(t)-U_1(t)$\\$\dot{U}_1(t)=-\frac{U_1(t)}{R_1(t)C_1(t)}+\frac{I(t)}{C_1(t)}$}}&{\makecell[l]{$U_{oc}(t, SOC)$, $R_1(t)$\\ $C_1(t)$, $R_0(t)$}}\\

{\makecell[l]{Dual polarization model \\
\begin{minipage}[b]{0.2\columnwidth}
		\centering
		\raisebox{-.5\height}{\includegraphics[width=\linewidth]{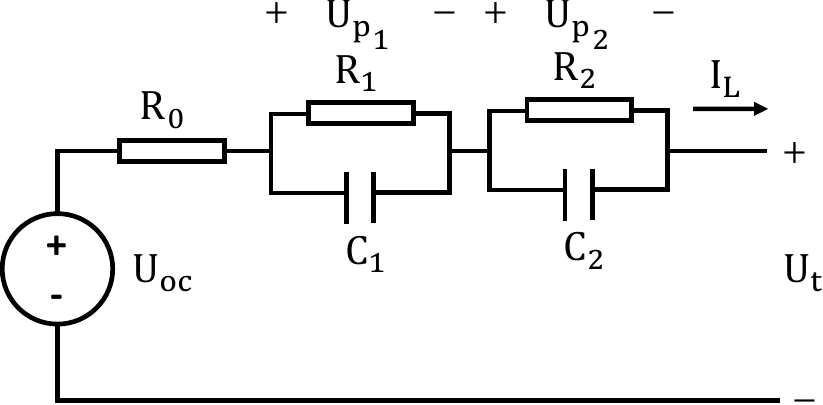}}
\end{minipage}}} &\cite{SDIEEECST2019,WLJES2020, YXWiley2021} & {\makecell[l]{$U_t(t) = U_{oc}(t, SOC) - I(t)R_0(t)$\\$-U_1(t)-U_2(t)$\\$\dot{U}_1(t)=-\frac{U_1(t)}{R_1(t)C_1(t)}+\frac{I(t)}{C_1(t)}$\\$\dot{U}_2(t)=-\frac{U_2(t)}{R_2(t)C_2(t)}+\frac{I(t)}{C_2(t)}$}}&{\makecell[l]{$U_{oc}(t, SOC)$, $R_1(t)$, \\$C_1(t)$, $R_2(t)$, \\$C_2(t)$, $R_0(t)$}}\\
\hline
\hline
\end{tabular}
\end{center}
\end{table*}
Although IOMs with RC pairs can enhance the modeling accuracy without excessively increasing computational burden, such models lack of physical significance, which hinders the exploration of inner mechanisms of LIBs \cite{XHENERGY2022}.

\subsubsection{Fractional-Order Models}

\begin{figure*}[!t]
\centering
\includegraphics[width=6in]{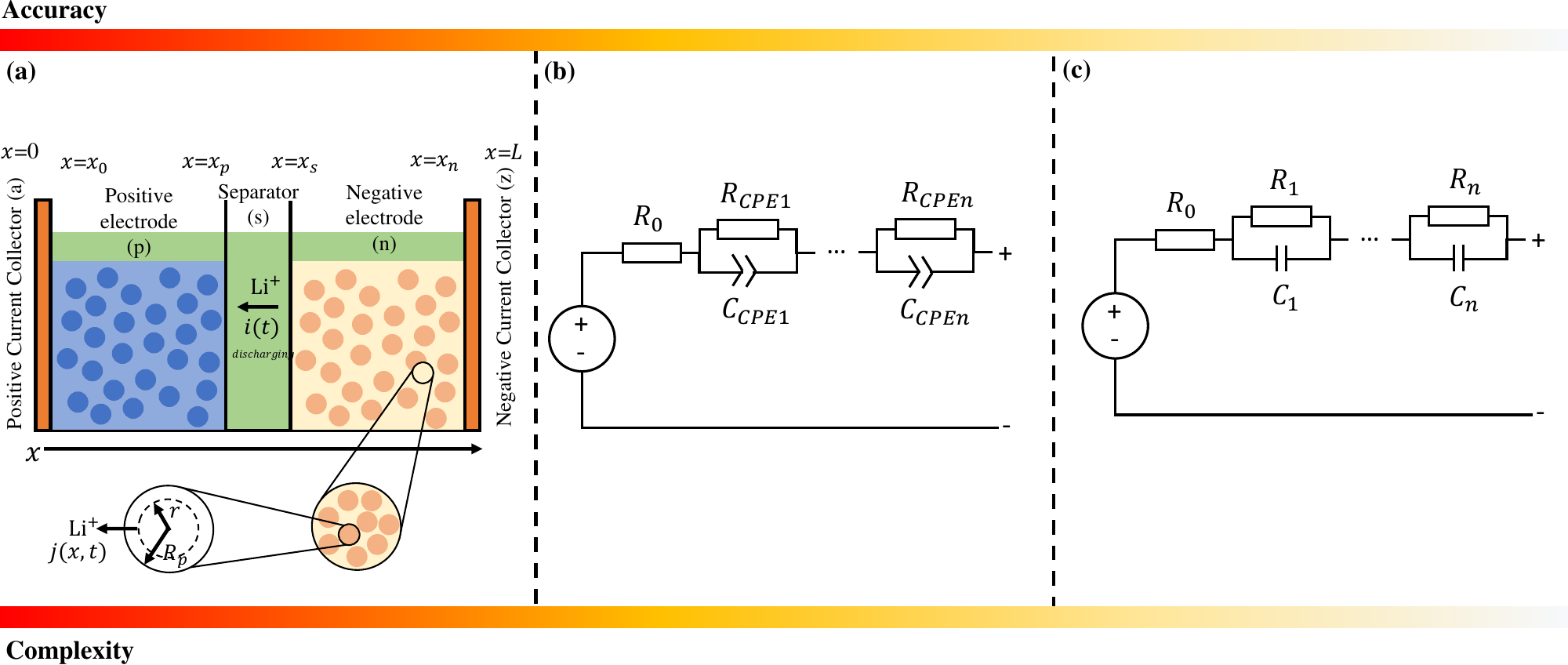}
\caption{Battery models applied for fault diagnosis (from left to right: (a) physics-based EM; (b) FOM; (c) IOM)}
\label{Fig1}
\end{figure*}

FOMs can be viewed as a type of improved ECMs by replacing the conventional capacitance in $RC$ pair with a constant phase element (CPE) or Warburg element to form a $RC_{CPE}$ or $RC_W$ pair. They can overcome the drawbacks of IOMs with finite RC pairs in accurately describing the frequency domain impedances of LIBs and thus significantly facilitate model interpretability to strong battery nonlinearities. As per the in-depth research on LIBs \cite{JTTIE2018, HRElectro2021, DGJES2020},
FOMs have been recognized with preferrable model accuracy over generic IOMs. For instance, Farmann and Sauer \cite{AFDSAE2018} compared a total of seven IOMs/FOMs with different numbers of $RC/RC_{CPE}$ pair (up to three) in their study. They showed that a $RC_{CPE}$ pair, describing a set of electrochemical processes (e.g., charge transfer and diffusion processes), had an equivalent accuracy to three conventional $RC$ pairs, and the FOM with three $RC_{CPE}$ pairs indicated the highest accuracy with an error of less than 20 m$\Omega$ in EIS fittings. In a similar research by Lai et al. \cite{XLJCP2020}, FOMs outperformed the IOMs with the same structures in battery state estimation over the whole battery operating range. By far, FOMs have been adopted in a wide range of applications, such as SOC \cite{RXJTWSTVT2018, QZEnergy2019, YXAMM2020},
State of energy \cite{XLJPS2017} and State of health (SOH) \cite{JTTIE2018, XHHYCZTVT2018} estimation, fault diagnosis \cite{RYRXJCP2018, PHXCAST2022}, and optimal charging \cite{YWGZCZTTE2022}. However, the complex structures and the nonlinear representations of FOMs affect their adaptability. The studies in \cite{ZYRHTPE2021, SMJES2023} tried to update the model parameters of FOMs online with some searching algorithms, while resulting in heavy computational burden on on-board BMSs and thus making them infeasible for practical applications. Some other studies built extensive linkages between model parameters and their dependencies (e.g., SOC and temperature) for online adaptions \cite{XTYWKYJPS2019}, where it required labor-intensive characterizations of model parameters over a wide battery operating range and storing such a mass of data remarkably reduced the usable memory of on-board BMSs. Therefore, it is of great challenge for a single FOM with limited adaptability to comprehensively capture the battery dynamics under various loads, SOCs, and temperatures.
Fig. \ref{Fig1} shows a summary of the above-mentioned models and an illustrative depiction of the performance characteristics of models from model accuracy and implementation complexity, which are two key metrics to select an appropriate model for model-based diagnosis methods.

\section{State-Space Representation}

Although EMs offer evident advantages over ECMs in terms of internal dynamics disclosure for LIBs, their high nonlinearity and numerous battery parameters make them difficult in real-time computation and parameter identification. Therefore, an ECM is the suitable choice in the model-based battery fault diagnosis methods. Regardless of IOMs or FOMs, a battery model can be generally represented as follows:
\begin{eqnarray}
\left\{\begin{array}{ll}
\frak{D}^\alpha x(t) &=A(x(t)) x(t) + B(x(t)) u(t) + \bar{B}(x(t)) \bar{f}(t) + w(t)\\
y(t) &= C(x(t)) x(t) + D(x(t)) u(t) + \bar{D}(x(t)) \bar{f}(t) + v(t),
\end{array}\right.\label{eq0}
\end{eqnarray}
where $x(t)$ denotes the real-time battery dynamic state vector, $\frak{D}^\alpha$ represents the integral-differential operator with respect to $t$,
$\alpha$ represents the intergral-differential order ($0<\alpha \leq 1$). When $\alpha=1$, \eqref{eq0} is in the integral form and can be used to represent the IOM; when $0<\alpha < 1$, \eqref{eq0} is in the fractional form and can be used to represent the FOM. $y(t)$ represents the measurable output of the model, $u(t)$ represents an input current, $\bar{f}(t)=\left[f(t), f_i(t), f_v(t)\right]^T$ represents the augmented vector of battery faults with $f_i(t)$ denoting a current sensor fault, $f_v(t)$ denoting a voltage sensor fault, $f(t)$ denoting a battery fault, which may affect the system state eqation and sensor measurement equation in a coordinated way, $w(t)$ represents the modeling errors, $v(t)$ represents the measurement errors. $A(x(t)), B(x(t)), C(x(t)), D(x(t))$ represent the real-valued time-varying battery parameter matrices, $\bar{B}(x(t)) = [B_f(x(t)), B(x(t)), 0] , \bar{D}(x(t)) = [D_f(x(t)), D(x(t)), 1]$ represent the real-valued time-varying fault configuration matrices. Thereinto, $B_f(x(t))$ and $D_f(x(t))$ are determined based on the physical configuration of the system, for example, by letting $B_f(x(t))=B(x(t))$ and $D_f(x(t))=D(x(t))$, the effect of battery SC fault can be investigated; by letting $B_f(x(t))=0$ and $D_f(x(t))=1$, the effect of battery connection faults can be explored.

\subsection{State Vectors}
The state vector $x(t)$ commonly involves a polarization voltage $U_1$ and a state of charge $SOC$ as $\left[U_1,~SOC\right]^T$ \cite{RXQYUTPE2019, MSTPE2021, ZYJES2021}
or includes a terminal voltage $U_t$ augmented as $\left[U_1,~SOC,~U_t\right]^T$ \cite{PSTVT2021}, 
\cite{JWTIE2020, XCTVT2016, XCJPS2014}.
For such a state vector, battery fault occurrence can be triggered based on the comparison between the estimated and reference state.
The fault diagnosis performance has certain limitations in fault response time and fault estimation accuracy.
To cope with restrictions,  the fault can be incorporated into a battery state (e.g., SC current, sensor fault) as $\left[U_1,~SOC,~f\right]^T$ \cite{QuanqingyuAE2022, YXTPE2022}. The fault severity can be directly estimated from the battery state, which leads to the improvement in fault response time and fault estimation accuracy.
A detailed comparative analysis of different state vectors is demonstrated in Table \ref{tab3}.

\begin{table*}[!htb]
\begin{center}
\caption{A comparative analysis of different state vectors}
\label{tab3}
\begin{tabular}{ l  l  l l  l}
\hline
\hline
State Vectors & References & Parameter matrices & Features & Limitations\\
\hline
$x(t)=\begin{bmatrix}U_1\\SOC\end{bmatrix}$ & {\makecell[l]{\cite{RXQYUTPE2019, MSTPE2021}\\\cite{ZYJES2021}}} & \makecell[l]{$A(x(t))=\begin{bmatrix}-\frac{1}{\tau(t)} & 0\\
0 & 0\end{bmatrix}$\\
$B(x(t))=\left[\frac{1}{C_1(t)}, -\frac{1}{C_n}\right]$\\
$C(x(t)) = \left[-1, 0\right]$\\ $D(x(t)) =-R_0(t)$,\\
$\tau(t) = R_1(t)C_1(t)$}  &\makecell[l]{Lowest rank \\for parameter\\ matices} & \makecell[l]{Difficulty in \\SC fault \\evaluation}\\
\hline
$x(t)=\begin{bmatrix}U_1\\SOC\\U_t\end{bmatrix}$ & {\makecell[l]{\cite{PSTVT2021, XYTPE2024, JWTIE2020}\\\cite{XCTVT2016, XCJPS2014}}} & \makecell[l]{$A(x(t))=\begin{bmatrix}-\frac{1}{\tau(t)} & 0 & 0\\
0 & 0 & 0\\
0 & \frac{\rho}{\tau(t)} & -\frac{1}{\tau(t)}\end{bmatrix}$\\
$B(x(t))=\left[\frac{1}{C_1(t)}, -\frac{1}{C_n}\right]$\\
$C(x(t)) = \left[0, 0, 1\right]$,\\
$\rho$ denoting the slope of OCV-SOC} &\makecell[l]{Linearization of \\parameter matrix\\ $C$} & \makecell[l]{Assumption for\\ $\dot{U}_t$ derivation}\\
\hline
 $x(t)=\begin{bmatrix}U_1\\SOC\\\bar{f}\end{bmatrix}$& \cite{QuanqingyuAE2022, YXTPE2022} & Determined by fault type &\makecell[l]{Co-estimation of\\ battery state\\ and fault} & \makecell[l]{Assumption for\\ $\dot{\bar{f}}$ derivation}\\
\hline
\hline
\end{tabular}
\end{center}
\end{table*}

\subsection{Parameter Matrices}

The system parameter matrices $A(x(t))$, $B(x(t))$, $C(x(t))$ and $D(x(t))$ implicitly involve the internal resistance $R_0$, polarization resistance $R_1$, polarization capacitance $C_1$ and the battery nominal capacity $C_n$. Since the accuracy of these battery parameters significantly affects the accuracy of state estimation, parameter identification plays an essential role in the model-based fault diagnosis method. It can be generally categorized into offline methods and online methods \cite{ZSSci2021}.

\subsubsection{Offline parameter identification}

\begin{figure}[!t]
\centering
\includegraphics[width=6in]{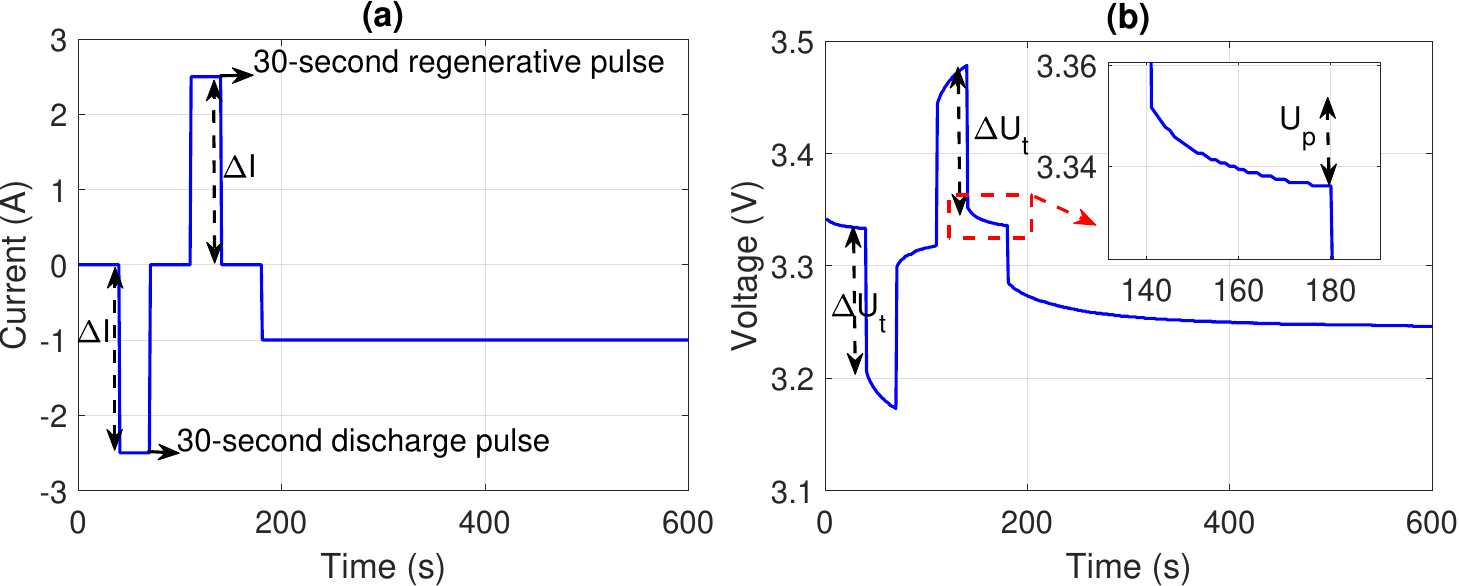}
\caption{HPPC test: (a) Current profile; (b) Voltage response}
\label{Fig2}
\end{figure}

The Hybrid Pulse Power Characterization Test can be adopted to identify the battery parameters \cite{ HZBatteries2022,JCUSA2015}. The objective of this test is to determine the 30-second discharge-pulse and the 30-second regenerative-pulse at each 10$\%$ increment relative to the operating capacity of a battery \cite{FZAE2016, SSJPS2013}. Between each pair of discharge and regenerative pulses, battery is discharged to the next 10$\%$ increment based on operating capacity. The pulse profile and the voltage response are respectively shown in Fig. \ref{Fig2}.

Considering that the measurement of voltage and current in a battery system is commonly performed at a specific sampling frequency $1/T_s$, the continuous time variables in terms of $t \in \mathbb{R}^+$ should be discretized at sampling instants of $t = k T_s$ with $k$ denoting the sampling index ($k \in \mathbb{Z}^+$) and $T_s$ denoting the sampling interval/period. Assuming that the immediate voltage jump occurs at time $\tilde{k}$, the internal resistance $R_{0, \tilde{k}}$ can be calculated by
\begin{align}
\label{eq1}
R_{0, \tilde{k}} =  \frac{\Delta U_t}{\Delta I}=
\frac{U_{t, \tilde{k}-1}-U_{t, \tilde{k}}}{I_{\tilde{k}-1}-I_{\tilde{k}}}.
\end{align}
While the polarization resistance $R_1$ and polarization capacitance $C_1$ can be estimated from the continuous voltage change after the voltage leaps by the least square algorithm \cite{ZCTIE2021, XPJP2022}. The formula that needs to fit can be given by 
\begin{align}
\label{eq2}
U_{1, k+1} = \text{exp}(\frac{-T_s}{R_{1, k}C_{1, k}})U_{1, k} + R_{1, k} I_{L, k} \left(1-\text{exp}(\frac{-T_s}{R_{1, k}C_{1, k}})\right).
\end{align}

\subsubsection{Online parameter identification}

Battery parameters feature slow-varying characteristics, and are jointly influenced by a series of factors (e.g., SOC, capacity, current and temperature). Online parameters identification offers more superiority in terms of adaptability, accuracy and computational efficiency for EV applications. It can be classified into recursive methods and non-recursive methods.

To apply recursive methods, the following three key points should be addressed, including model discretization, implementation assumption and algorithm selection.

For the model discretization, according to the continuous-time domain electrical dynamic derived from the circuit theory, we can obtain the battery behaviour in the frequency domain via Laplace transform. After substituting $s = \frac{1-z^{-1}}{1+z^{-1}}$, the discrete battery model can be reconstructed into the following form:
\begin{align}
\label{eq3}
U_{t, k} = x_k \theta_k + v_k,
\end{align}
where $\theta_k$ denotes the parameter vector to be estimated at time step $k$ (or the sampling time $kT_s$), $x_k$ denotes the information vector at time step $k$, and $v_k$ represents the unknown measurement noise. More details of the discretization for a first-order equivalent circuit model can be found in \cite{YXTTE2023}. 

Then, certain implementation assumptions need to be made in advance. For instance, considering slow variation of model parameters, it is assumed that the parameters remain unchanged within adjacent time intervals \cite{XDTIE2022}. Thus, we can get $\theta_{k+1} = \theta_k+w_k$, where $w_k$ denotes the unknown modeling error and $\theta_k$ involves $U_{oc}$, $R_1$, $R_0$ and $C_1$ dependent on the discretization model.

The next step is to select the appropriate identification algorithms, among which the recursive least-square (RLS) algorithm is commonly used for online identification due to its strong adaptability and low computational efforts. For example, 
Tan et al. \cite{XTJPS2021} iteratively updated the model parameters for minimizing the sum of squared errors between the estimated and measured terminal voltages.
To alleviate data saturation and slow convergence speed in the RLS algorithm, a forgetting factor is commonly introduced to give different weights to the old and recent data \cite{KSTIS2010}. The recent data are assigned more weight than the old data, thereby effectively reducing the burden of old data on the calculation.
In \cite{NSJES2022}, a forgetting factor was employed to prioritize the recent data, which realized adaptive track of the parameter changes over time.
In \cite{GPJPS2006}, Kalman filter (KF) was chosen to deal with noises abiding by Gaussian's distribution.
It demonstrates an advantage in realizing co-estimation for battery state and model parameters.
Although the above-mentioned recursive methods have been proven to be effective, the issue to find out the globally optimal parameters hinders their identification accuracy.

Non-recursive methods, on the other hand, take the absolute error between the measured and estimated terminal voltage throughout the dataset as the fitness function. They can reduce the probability of being trapped at a local minimum to approach the global optimal solution continuously and have the potential to achieve higher accuracy than recursive methods.
In \cite{RYJCR2018, RYAE2017}, Genetic algorithm (GA) was employed to optimize the model parameters and the objective function $F$ is set to be the sum of the least square error between the estimated and measured terminal voltage, namely
\begin{align}
\label{eq4}
F = \text{min}\left\{\sum^L_{k=1}(U_{t, k}-\hat{U}_{t, k})^2\right\},
\end{align}
with $L$ being the size of measurement. 
Rahman et al. \cite{MRJPS2016} applied the particle swarm optimization (PSO) to identify electrochemical model parameters under operating conditions and defined a cost function to minimize the error between estimated and measured terminal voltage.
Although non-recursive methods exhibit high efficiency in searching for global optimal solutions, there exists one limitation that impacts their online implementation, namely the execution of these non-recursive methods will inevitably increase computation complexity, which may prolong the algorithm's runtime and subsequently delay fault diagnosis. A comparison among some existing parameter identification methods is performed and listed in Table \ref{tab4}.

\begin{table*}[!t]
\begin{center}
\caption{A comparison among some existing parameter identification methods}
\label{tab4}
\begin{tabular}{l l l l l}
\hline
\hline
 & & References &  Advantages & Disadvantages\\
\hline
 {Offline}& HPPC+OCV & {\makecell[l]{\cite{HZBatteries2022, JCUSA2015, JCUSA2015}\\ \cite{FZAE2016, SSJPS2013, XPJP2022}}}&  {\makecell[l]{$\cdot$ Simple implementation}} & {\makecell[l]{$\cdot$ Real-time capability deficiency\\ $\cdot$ Dedicated experiments}}\\
\hline
 {\multirow{4}*{Online}}& Recursive & {\makecell[l]{\cite{XTJPS2021, NSJES2022}\\ \cite{GPJPS2006, QJENERG2018}}} & {\makecell[l]{$\cdot$ Low computational cost\\ $\cdot$ Moderate precision}} & {\makecell[l]{$\cdot$ Local minimum entrapment\\ $\cdot$ Model linearization \\$\cdot$ Historical data influence}}\\
& Non-recursive&{\makecell[l]{\cite{RYJCR2018, RYAE2017}\\ \cite{MRJPS2016}}} & {{$\cdot$ High precision}}&{\makecell[l]{$\cdot$ Long identification time\\ $\cdot$  High computation cost}}\\
\hline
\hline
\end{tabular}
\end{center}
\end{table*}

\section{Fault Mechanisms}

LIBs suffer from potential safety issues in practice inherent to their energy-dense chemistry and flammable materials. From the perspective of electrical faults, fault modes can be divided into battery faults and sensor faults.

\subsection{Battery Faults}
\begin{figure*}[!t]
\centering
\includegraphics[width=6in]{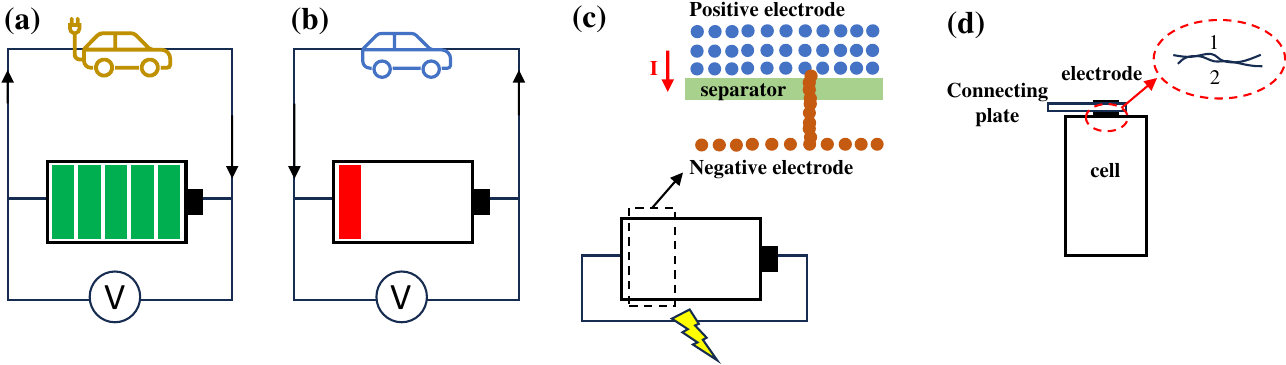}
\caption{Battery faults: (a) overcharge; (b) overdischarge; (c) SC fault; (d) connecting fault}
\label{Fig3}
\end{figure*}

Battery faults mainly include overcharge/overdischarge fault, battery connection fault, and SC fault, as illustrated in Fig. \ref{Fig3}.

\subsubsection{Overcharge/overdischarge fault}
Overcharge/overdischarge refers to the behavior of continuing to charge/discharge the battery even when the battery reaches the upper/lower cutoff voltage. The triggering of the overcharge/overdischarge fault comes from the malfunction of a charger and the failure of sensors. The repeated overcharge/overdischarge is often accompanied by the losses of cyclable Li ions and active material, which will accelerate battery degradation \cite{DOJES2022, JXJES2022}. Furthermore, overcharging may lead to Li deposition and contribute to an irreversible phase change and even the collapse of the cathode structure with a gas release and heat generation \cite{QYElectro2015}.
Overdischarging can cause the breakage of electrode materials, decomposition of the solid electrolyte interface and dissolution of copper, resulting in capacity reduction and cathode and anode structure deterioration\cite{XLEActa2018}.

\subsubsection{Battery connection fault}

To meet voltage and energy demands, LIBs are connected in series or parallel to compose a battery pack. During EV operation, vibrations may lead to loose or poor electrical connections between battery cells in the pack \cite{MMEnergy2018}. The resultant abnormality in the intercell contact resistance is defined as battery connection fault \cite{YKAE2020, DSEnergy2023}. Such a type of fault can cause an uneven current flow into a cell, leading to a severe cell imbalance in a battery pack and an increase in heat generation\cite{MMTVT2020}.

\subsubsection{SC faults}

SC faults can be classified into ISC faults and ESC faults.
An ISC happens when the insulating separator between the electrodes fails. The failure of the separator can be attributed to melting due to high temperature, cell deformation, the formation of dendrite, or compressive shock. It can result in contact between the anode and the cathode \cite{XFAE2016} and the formation of an internal current loop within a battery, leading to continuous discharge, heat accumulation and a high risk of thermal runaway for a battery \cite{MSJES2022}.
On the other hand, an ESC occurs when the positive and negative terminals make contact externally\cite{ZCAE2021}.
It can be caused by the deformation during a car collision, water immersion or contamination with conductors \cite{RYJEE2021, ZAAE2023}.
Although ESCs usually cause relatively lower temperature rise than ISCs, they can still result in battery damage and other serious consequences (e.g. electrolyte leakage, fire and even explosion)\cite{ZCBZ}.

No matter of ISC or ESC, an SC fault can be electrically equivalent to a resistance short-circuiting the battery terminal at different magnitudes. Based on the magnitude of SC resistance, SC faults can also be categorized into hard SC faults and soft SC faults. The threshold utilized for differentiating hard and soft SC faults in the existing literature are summarized in Table \ref{tab5}.

\begin{table}[!htb]
\begin{center}
\caption{Typical thresholds for hard and soft SC faults in the literature}
\label{tab5}
\begin{tabular}{ l  l}
\hline
\hline
References & Thresholds\\
\hline
\cite{XLESM2021, MSJES2022, JMJES2020} & 10~$\Omega$\\
\cite{SBJPS2022, SBIscience2023, XKJES2020},  & 5.5~$\Omega$ (C/3.7~A)\\
\hline
\hline
\end{tabular}
\end{center}
\end{table}

\subsection{Sensor Faults}

The sensors discussed hereinafter mainly include voltage sensors and current sensors that are used to monitor voltage and current of a battery. A direct impact of sensor faults is that BMS cannot obtain the accurate working status of a battery and send out the wrong control signals, leading to the unconscious abusive operation on a battery system \cite{ZLEnergies2015}. For instance, when a voltage sensor fault occurs, the overcharge/overdischarge protection based on the cutoff voltage thresholds may fail and also may mislead a BMS to aggravate the system inconsistency; when a current sensor fault occurs, BMS may misjudge the working state of a battery and give incorrect control commands, causing the accelerated battery aging \cite{ZLAE2017, BXJPS2017}.

Recalling $\bar{f}(t)=\left[f(t), f_i(t), f_v(t)\right]^T$ in the battery model \eqref{eq0}, the voltage sensor fault $f_v(t)$ and current sensor fault $f_i(t)$ can be specifically described as:
\begin{align}
\label{eq5}
&f_i(t) = a s(t)  + b I(t),\\
&f_v(t) = c s(t) +d U_t(t),
\end{align}
where $s(t)$ denotes the arbitrary bias fault injection signal (e.g., step function, periodic function, intermittent function), $U_t(t)$ denotes the measured terminal voltage and $I(t)$ denotes input current. The above sensor fault models can accommodate a variety of sensor faults by prescribing the values of $a$, $b$, $c$ and $d$. Table \ref{tab6} summarizes different some typical sensor faults under different values of $a$, $b$, $c$ and $d$.

\begin{table}[!htb]
\begin{center}
\caption{A summary of different sensor faults under different values of $a$, $b$, $c$ and $d$}
\label{tab6}
\begin{tabular}{ l  l  l  l l l}
\hline
\hline
Sensor source & Fault type & $a$ & $b$ & $c$ & $d$\\
\hline
Voltage sensor & Bias& 0 & 0 & 1 & 0\\
Voltage sensor & Gain& 0 & 0 & 0 & 1\\
Voltage sensor & Intermittent& 0 & 0 & 1 & 1\\
Current sensor & Bias& 1 & 0 & 0 & 0\\
Current sensor & Gain& 0 & 1 & 0 & 0\\
Current sensor & Intermittent& 1 & 1 & 0 & 0\\
\hline
\hline
\end{tabular}
\end{center}
\end{table}

\section{Modeling Uncertainties}

An accurate dynamic model of a process or target can be barely established. The existence of unknown modeling uncertainties in \eqref{eq0} inevitably renders the accurate estimation of the real-time battery state $x(t)$ difficult or challenging. The following different sources of uncertainties are briefly outlined for a battery state-space model.

\subsection{Modeling Errors and Measurement Noises}

The parameter matrices in a battery state-space model are obtained through offline or online identification methods. It is evident that there may be discrepancies between the estimated parameters and actual parameters, which can cause modeling errors. When referring to measurement noises, the measurement of the terminal voltage and input current are inevitably affected by measurement accuracy limitations coming from sensors.

To deal with modeling uncertainties arisng from modeling errors and measurement noises, two unknown nonlinear terms $w(t)$ and $v(t)$ can be introduced in the state-space equations \eqref{eq0}. Then some existing modeling methods can be classified into two types: stochastic methods and deterministic methods. 

The stochastic methods typically impose some a priori assumptions on the statistical properties of modeling uncertainties. An emphasis is then laid on minimizing the variance of the state estimation error between the estimated state and true state. For example, the authors in \cite{YJMEASUREMENT2019, CWIJES2016} assumed a Gaussian probability distribution of measurement noises and then applied KF-based approaches for estimation.
The deterministic methods, on the other hand, require the assumption of the hard bounds of modeling uncertainties which are either energy-bounded or amplitude-bounded.
The core idea of such methods comes to suppress the influence of modeling uncertainties on state estimation. Zhu et al. \cite{QZTVT2017} designed an $H_\infty$ observer based on an energy-bounded modeling uncertainties.
Ouyang et al. \cite{QOuTPE2020} assumed the model parameter estimation remains bounded and developed a robust observer to suppress the influence coming from modeling uncertainties.

In addition to the impact on the accuracy of state estimation, modeling uncertainties have an influence on the residual evaluation-based fault diagnosis methods. Theoretically, in a fault-free battery system, the residual signal between the estimation and measurement is expected to be zero. However, in practical applications, the residual tends to fluctuate around zero. Therefore, rather than solely comparing with zero, it is necessary to establish a threshold for fault diagnosis. For example, the 3$\sigma$ rule is commonly applied to set the threshold for determining outliers. Zhao et al. \cite{YZAE2017} detected the abnormal changes of battery terminal voltages according to 3$\sigma$ multi-level screening strategy.  Lin et al. \cite{MLIEEE2022} calculated the failure threshold by combining the 3$\sigma$ rule and multiscale permutation entropy of batteries. In an empirical context, the Monte-Carlo simulation can be employed to identify the fault-free range for different battery types or different drive cycle conditions prior to being utilized for fault diagnosis. Dey et al. \cite{SDTCT2019} found the residual probability distribution under normal conditions by collecting data from Monte Carlo simulations.

\subsection{Aging Effects}

In a battery cycling process, the inevitable side reactions can cause the loss of lithium inventory and the loss of active materials, leading to a decrease in battery capacity and an increase in internal resistance \cite{JYEnergy2023,JYYCEnergy2022,JZYWNC2022}.

Aside from the modeling errors and measurement noises, the capacity estimation error also impact the performance of battery fault diagnosis. As demonstrated below, 
\begin{align}
\label{eq6}
\dot{SOC}(t) = \frac{I(t)}{C_n(t)},
\end{align}
it becomes apparent that the accuracy of SOC estimation heavily relies on the identified battery capacity. To address errors stemming from the aging effects, the concurrent estimation of battery capacity and battery state has the potential to enhance the accuracy of state estimation and fault diagnosis \cite{MCTPE2017}. Zou et al. \cite{YZJPS2015} devised two extended KFs (EKFs) with different time scales for combined SOC/SOH monitoring. The SOC was estimated in real-time by using a second-order EKF, and the SOH was updated offline in a fourth-order EKF which was triggered by the instantaneous voltage difference from the measured value.

\subsection{Measurement Outliers}

The measurement outliers are referred to as the anomalous measurements which significantly deviate from the normal values \cite{SSJPS2021}. Different from the above-mentioned modeling errors and measurement noises, it is uncommon to encounter outliers in the lab environment. For example, Chen et al. \cite{HCTII2023} put forward an online-outlier-detection method to detect and diagnose the type of outliers by means of the chi-square test. Combined with the EKF algorithm and Holt's two-parameter linear exponential smoothing method, an outlier-resistant KF algorithm is proposed to prevent the outlier-induced effect from degrading the estimation performance. 
In cloud-enabled applications, the transmission of measurement data to the cloud server for recordkeeping or the measurements in the scenarios involving sudden acceleration or abrupt braking in real EV applications are often vulnerable and can incur measurement outliers.
Given the majority of the existing model-based estimation and diagnosis methods rely on voltage measurements, the presence of measurement outliers can result in a complete failure of battery state estimation and fault diagnosis \cite{JMPro2010}. 

\section{State Observers and Algorithm Implementation}

State estimation and prediction of a battery system are the basis for a model-based fault diagnosis algorithm.
In particular, to achieve state estimation and prediction, a suitable state observer is always needed.
Based on whether the observer gains are computed through online iteration or offline calculation, the existing state observers can be roughly categorized into two groups: online state observers and offline state observers.

\subsection{Online State Observers}
KFs and its variants are typical online observers and broadly used in model-based fault diagnosis algorithms to estimate battery states, such as SOC and terminal voltage. More specifically, a KF operates in a recursive manner involving two main steps: (1) predicting the system state and output, and (2) updating the system state based on the output error. It should be noted that a KF is often designed for linear dynamic systems. However, when the KF is applied to a nonlinear battery system, certain modifications are necessary for efficient implementation. To linearize the nonlinear OCV-SOC curve, Plett et al. \cite{GPJPS2004} performed SOC estimation by using EKF, which expands the nonlinear function with partial derivatives based on the Taylor expansion. In \cite{RXTVT2011}, an adaptive EKF was designed for LIB SOC estimation via an improved thevenin model.
The implementation procedure of EKF is shown in three steps: initialization, time update and measurement update, as summarized in Algorithm \ref{algo1}.

\begin{algorithm}
    \caption{Implementation of EKF}
\label{algo1}
\footnotesize

\textbf{Step 1 Battery model discretization:}

In the absence of faults, the IOM-based battery system \eqref{eq0} can be discretized at the sample rate of $T_s$ in a linear form as:
\begin{align}
 \nonumber x_k &= f(x_{k-1}, u_{k}, \omega_{k-1}) \approx A_{k-1}x_{k-1} +  B_{k-1}u_{k}+\omega_{k-1},\\
 \nonumber y_k &= h(x_k, u_k, v_k) \approx C_kx_k + D_ku_k +v_k,
\end{align}
where $x_k$ denotes the battery state, $y_k$ represents the measurable output, $A_{k-1}$, $B_{k-1}$, $C_k$ and $D_k$ are battery-related parameters that can be obtained in real time. $\omega_{k-1}$ and $v_k$ are independent, zero-mean, Gaussian modeling errors and measurement noises of covariance matrices $\Sigma_\omega$ and $\Sigma_v$\;

\textbf{Step 2 Online implementation:}

\textit{\textbf{Initialization:}}

\textit{Set} $\hat{x}_0^+=\mathbb{E}\left[x_0\right]$, $\Sigma_{\tilde{x}, 0}^+=\mathbb{E}\left[\left(x_0-\hat{x}_0^+\right)\left(x_0-\hat{x}_0^+\right)^T\right]$, maximal simulation time $T_{max}$, sampling time $T_{s}$\;

\textbf{For} $k = 1:T_s:T_{max}$

\textit{\textbf{Time update:}}

\textit{Collect} input current $u_k = I_{k}$ and \textit{collect} battery-related parameters  $A_{k-1}$, $B_{k-1}$, $C_k$, $D_k$\;

\textit{Compute} battery state: $\hat{x}_k^{-} = A_{k-1}\hat{x}_{k-1}^+ +  B_{k-1}u_{k}$\;

\textit{Compute} covariance matrix of state error: $\Sigma_{\tilde{x}, k}^- = A_{k-1}\Sigma_{\tilde{x}, k-1}^+A_{k-1}^T+\Sigma_\omega$\;

\textit{\textbf{Measurement update}} \cite{GPJPS2004}\textit{\textbf{:}}

\textit{Collect} terminal voltage $y_k=U_{t, k}$\;

\textit{Calculate} innovation: $e_k = y_k-(C_k\hat{x}_{k}^-+D_ku_{k})$\;

\textit{Compute} gain matrix: $K_k = \Sigma_{\tilde{x}, k}^-C_{k}^T\left[C_k\Sigma_{\tilde{x}, k}^-C_k^T+\Sigma_v\right]^{-1}$\;

\textit{Correct} state estimation: $\hat{x}_k^+ = \hat{x}_k^-+K_ke_k$\;

\textit{Compute} covariance matrix: $\Sigma_{\tilde{x}, k}^+= \left(I-K_kC_k\right)\Sigma_{\tilde{x}, k}^-$\;

\textbf{End}
\end{algorithm}

The Sigma-point Kalman filter (SPKF) represents an alternative to generalize the KFs to estimate nonlinear system states \cite{GLPLETTJPS2006}. The detailed procedure is illustrated in Algorithm \ref{algo2}.

\begin{algorithm}
    \caption{Implementation of SPKF}
\label{algo2}
\footnotesize
\textbf{Step 1 Battery model discretization:}

In the absence of faults, the IOM-based battery system \eqref{eq0} can be discretized at the sample rate of $T_s$ in a nonlinear form as:
\begin{align}
\nonumber x_k &= f(x_{k-1}, u_{k}, \omega_{k-1}),\\
\nonumber y_k &= h(x_k, u_k, v_k),
\end{align}
where $x_k$ denotes the battery state, $y_k$ represents the measurable output, $\omega_{k-1}$ and $v_k$ are independent, zero-mean, Gaussian modeling errors and measurement noises of covariance matrices $\Sigma_\omega$ and $\Sigma_v$\;

\textbf{Step 2 Online implementation:}

\textit{\textbf{Definition:}}

\textit{Let} $x_k^a=\left[x_k^T, \omega_k^T, v_k^T\right]^T$, $\mathcal{X}_k^a=\left[(\mathcal{X}_k^x)^T, (\mathcal{X}_k^\omega)^T, (\mathcal{X}_k^v)^T\right]^T$\;

 $p$ is $2 \times$ dimension of $x_k^a$, $L$ is the dimension of $x_k$\;

\textit{\textbf{Initialization:}}

\textit{Set} $\hat{x}_0^+ = \mathbb{E}\left[x_0\right]$, $\hat{x}_0^{a, +} = \mathbb{E}\left[{x}_0^a\right]$, $\Sigma_{\tilde{x}, 0}^+ = \mathbb{E}(x_0-\hat{x}_0^+)(x_0-\hat{x}_0^+)^T, \Sigma_{\tilde{x}, 0}^{a, +} = \mathbb{E}(x_0^a-\hat{x}_0^{a, +})(x_0^a-\hat{x}_0^{a, +})^T$, maximal simulation time $T_{max}$, sampling time $T_{s}$\;

\textbf{For} $k = 1:T_s:T_{max}$

\textit{\textbf{Time update:}}

\textit{Collect} input current $u_k = I_{k}$\;

\textit{Compute} battery state: $\chi_{k-1}^{a, +} = \left[\hat{x}_{k-1}^{a, +}, \hat{x}_{k-1}^{a, +}+\gamma\sqrt{\Sigma^{a, +}_{\tilde{x}, k-1}}, \hat{x}_{k-1}^{a, +}-\gamma\sqrt{\Sigma^{a, +}_{\tilde{x}, k-1}}\right]$, $\chi_{k ,i}^{x, -} = f\left(\mathcal{X}_{k-1, i}^{x, +}, u_{k}, \mathcal{X}_{k-1, i}^{\omega, +}\right)$, $\hat{x}_k^-=\Sigma_{i=0}^p \alpha_i^{(m)}\mathcal{X}_{k, i}^{x, -}$ with $\gamma  = \sqrt{L+\lambda}$,  $\alpha_0^{(m)} = \frac{\lambda}{L+\lambda}$, $\alpha^{(m)}_{i, i=1,\cdots, p} = \frac{1}{2(L+\lambda)}$, $\lambda=\alpha^2(L+\theta)-L$, $\alpha$ being a positive proportional scaling factor with a range of $0-1$ and $\kappa$ is either 0 or $3-L$\;

\textit{Compute} covariance matrix of state error: $\Sigma_{\tilde{x}, k}^-=\Sigma_{i=0}^p\alpha_i^{(c)}\left(\mathcal{X}_{k, i}^{x, -}-\hat{x}_k^-\right)\left(\mathcal{X}_{k, i}^{x, -}-\hat{x}_k^-\right)^T$ with $\alpha_0^{(c)} = \frac{\lambda}{L+\lambda}+(1-\alpha^2+\beta)$ and $\alpha^{(c)}_{i, i=1,\cdots, p} = \frac{1}{2(L+\lambda)}$\;

\textit{\textbf{Measurement update}} \cite{GLPLETTJPS2006}\textit{\textbf{:}}

\textit{Collect} terminal voltage $y_k=U_{t, k}$\;

\textit{Calculate} output estimation: $\mathcal{Y}_{k, i}=h(\mathcal{X}_{k, i}^{x, -}, u_{k}, \mathcal{X}_{k, i}^{v, +})$, $\hat{y}_k = \Sigma_{i=0}^p \alpha_i^{(m)}\mathcal{Y}_{k, i}$\;

\textit{Compute} gain matrix: $\Sigma_{\tilde{y}, k}=\Sigma^p_{i=0} \alpha_i^{(c)}\left(\mathcal{Y}_{k, i}-\hat{y}_k\right)\left(\mathcal{Y}_{k, i}-\hat{y}_k\right)^T$, $\Sigma_{\tilde{x}\tilde{y}, k}^-=\Sigma^p_{i=0} \alpha_i^{(c)}\left(\mathcal{X}_{k, i}^{x, -}-\hat{x}_k^{-}\right)\left(\mathcal{Y}_{k, i}-\hat{y}_k\right)^T$, $L_k =\Sigma^{-}_{\tilde{x}\tilde{y}, k}\Sigma^{-1}_{\tilde{y}, k}$\;

\textit{Correct} state estimation: $\hat{x}_k^+ = \hat{x}_k^-+L_k\left(y_k-\hat{y}_k\right)$\;

\textit{Compute} covariance matrix: $\Sigma_{\tilde{x}, k}^+=\Sigma_{\tilde{x}, k}^--L_k\Sigma_{\tilde{y}, k}L_k^T$\;

\textbf{End}
\end{algorithm}

\begin{figure*}[!b]
\centering
\includegraphics[width=6in]{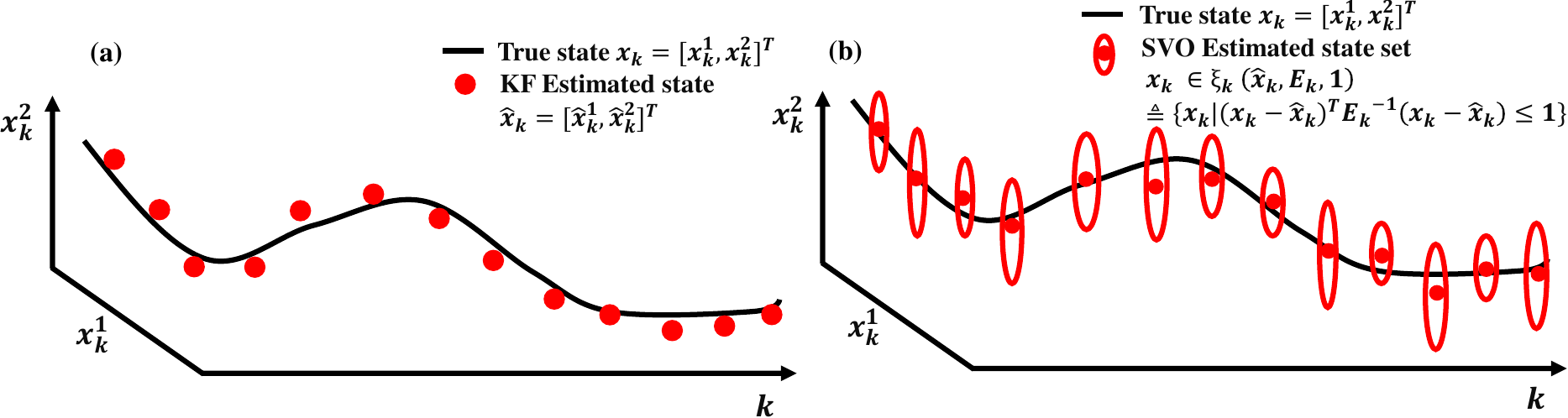}
\caption{Illustration of pointwise KF estimates and ellipsoidal set-valued estimates of the unknown true system state $x_k = [x_k^1, x_k^2]^T$
}
\label{Fig4}
\end{figure*}

Instead of calculating a pointwise estimation of the unavailable battery state through KF or its variants, Xu et al. \cite{YXTVT2023} determined an ellipsoidal estimation set via a set-valued observer (SVO) such that the actual battery states can be guaranteed to be enclosed in a set regardless of unknown but bounded modeling uncertainties. 
A salient difference between the KF and SVO is that the KF only computes 
a pointwise estimate (a single vector) of the signal of interest at each instant of time, while the SVO provides a bounding ellipsoidal set enclosing all possible state estimates in the state space due to unknown but bounded errors or noises. In other words, the KF-type pointwise observers cannot guarantee that the unknown signal of interest (due to unknown errors or noises) is always included in
some reliable confidence region at each instant of time \cite{GeX19TC1,GeX20INS}. 
Fig. \ref{Fig4} demonstrates the distinction of state estimates between the pointwise KF and the ellipsoid-based SVO at each instant of time $k$.

\begin{algorithm}
    \caption{Implementation of SVO }
\label{algo7}
\footnotesize
\textbf{Step 1 Battery model discretization:}

In the absence of faults, the IOM-based battery system \eqref{eq0} can be discretized at the sample rate of $T_s$ in a linear form as:

\begin{align}
\nonumber x_{k+1} &= f(x_{k}, u_{k+1}, \omega_{k}) \approx A_{k}x_{k} +  B_{k}u_{k+1} + \omega_{k},\\
\nonumber y_{k+1} &= h(x_{k+1}, u_{k+1}, v_{k+1}) \approx C_{k+1} x_{k+1} + D_{k+1}u_{k+1}+v_{k+1},
\end{align}
where $x_k$ denotes the battery state, $y_k$ represents the measurable output, $\omega_{k}$ and $v_{k+1}$ denote unknown but bounded modeling errors and measurement noises and confine to the following ellipsoids:
\begin{align}
\nonumber \omega_{k} &\in \xi\{0, Q_{k}, 1\},\\
\nonumber v_{k+1} &\in \xi\{0, R_{k+1}, 1\},
\end{align}
with $Q_{k}$, and $R_{k+1}$ being the given shape matrices for the elliposidal sets, $A_{k}$, $B_{k}$, $C_{k+1}$ and $D_{k+1}$ are battery-related parameters that can be obtained in real time\;

\textbf{Step 2 Online implementation:}

\textit{\textbf{Initialization:}}

\textit{set} maximal simulation time $T_{max}$, sampling time $T_{s}$\; 

\textbf{For} $k = 0:T_s:T_{max}$

\textit{Collect} battery-related parameters $B_k$, $C_{k+1}$ and $D_{k+1}$\;  

\textit{\textbf{Observer design:}}
\begin{align}
\label{eq9a}
 \hat{x}_{k+1} &= G_{k}\hat{x}_{k} +  B_{k}u_{k+1} + L_{k}(y_{k}-\hat{y}_{k}),\\
\label{eq10a} 
\hat{y}_{k+1} &= C_{k+1} \hat{x}_{k+1} + D_{k+1} u_{k+1}.
\end{align}

The observer gain matrices $L_k$ and $G_k$ can be determined via the following optimization problem \cite{YXTVT2023}
\begin{align}
&\nonumber \text{min~trace}(P_{k+1})\\
&\nonumber \text{s.t.} \Pi_{k}=\begin{bmatrix}
-P_{k+1} & \tilde{\Pi}_k\\
* & \tilde{\lambda}_k
\end{bmatrix} <0
\end{align}
where $P_{k+1}>0$ is a matrix sequence and  $\theta_{m,k}>0$ are scalar sequences to be solved, and
\begin{align}
\nonumber &\tilde{\Pi}_{k+1}=\left[(A_{k}-G_{k})\hat{x}_{k}, A_{k}E_{k}-L_kC_kE_k, I, -L_{k}D_{k}\right],\\
\nonumber &\tilde{\lambda}_k = \text{diag}(\sum_{m=1}^3 \theta_{m,k}-1, -\theta_{1,k}I, -\theta_{2,k}Q_{k-1}^{-1},-\theta_{3,k}R_k^{-1})
\end{align}
with $E_{k}$ coming from the factorization $P_{k} = E_{k}E_{k}^T$\;

\textit{\textbf{State estimation:}}

\textit{Collect} terminal voltage $y_k=U_{t, k}$ and input current $u_{k+1} = I_{k+1}$\;

\textit{Obtain} gain matrices $L_k$, $G_k$ from \textit{Observer design} and $E_{k+1}$ by solving the above optimization problem\;

\textit{Compute} estimate center $\hat{x}_{k+1}$ via \eqref{eq9a}, estimate output $\hat{y}_{k+1}$ via \eqref{eq10a}\;

\textit{Derive} estimate set $\xi_{k+1}(\hat{x}_{k+1}, E_{k+1}, 1)$ $\triangleq$ $\{x_{k+1}|(x_{k+1}-\hat{x}_{k+1})^TE_{k+1}^{-1}(x_{k+1}-\hat{x}_{k+1}) \leq 1\}$ based on $\hat{x}_{k+1}$ and $P_{k+1}$\; 
\textbf{End}
\end{algorithm}

\subsection{Offline State Observers}

So far, several different offline observers have been developed for achieving state estimation, such as Luenberger observers (LOs) \cite{XHFSYZ2010}, sliding mode observers (SMOs) \cite{BNEnegy2018, XCWSTVT2016}, proportional-integral observers (PIOs) \cite{JXTVT2014}, $H_\infty$ observers (HIOs) \cite{JXJPE2016}, etc., outlined in Algorithm \ref{algo3}-\ref{algo6}, respectively. It should be noted that each type of observer has its unique characteristics and may offer relative advantages when tackling specific estimation problems. For example, leveraging the sliding mode variable structure, SMOs drive the dynamic responses of the system continuously and purposefully to follow the preset trajectory function and ultimately converge to the predefined sliding mode surface to reconstruct the system states. HIOs are capable to ensure robustness against inaccurate initial system state and modeling uncertainties by introducing a performance index to suppress their influences on the state estimation.
PIOs differ from other types of observers primarily in its incorporation of an integrator element. 
The addition of such an integrator element confers to the observer more robustness with respect to uncertainties. 
Table \ref{tab7} presents a comparative analysis of the observers aforementioned with their main features and limitations being highlighted.

\begin{algorithm}
    \caption{Implementation of LO}
\label{algo3}
\footnotesize
\textbf{Step 1 Battery model discretization:}

In the absence of faults, the IOM-based battery system \eqref{eq0} can be expressed in a simplified continuous linear form as:
\begin{align}
\nonumber \dot{x}(t) &= f(x(t), u(t), \omega(t)) \approx Ax(t) +  Bu(t),\\
\nonumber y(t) & = h(x(t), u(t), v(t)) \approx C x(t) + Du(t),
\end{align}
where $x_k$ denotes the battery state, $y_k$ represents the measurable output, $A$, $B$, $C$ and $D$ represent constant battery parameter matrices\;

\textbf{Step 2 Offline observer design:}

\textit{Collect} constant battery parameter matrices $A$ and $C$\;

\begin{align}
\label{eq11a}
 \dot{\hat{x}}(t) &= A{\hat{x}}(t) +  Bu(t) + L\left(y(t)-{\hat{y}}(t)\right),\\
\label{eq12a}
 {\hat{y}}(t) &= C {\hat{x}}(t) + D u(t),
\end{align}
where the observer gain matrix $L$ needs to be determined such that $A-LC$ is Hurwitz stable \cite{XHFSYZ2010}\;

\textbf{Step 3 Online implementation:}

\textit{Set} maximal simulation time $T_{max}$\;

\textit{Obtain} gain $L$ from \textbf{Step 2}\;

\textbf{While} $t \leq T_{max}$

\textit{Collect} terminal voltage $y(t)=U_{t}(t)$, input current $u(t) = I(t)$, constant battery parameter matrices $B$ and $D$\;

\textit{Compute} state estimation $\hat{x}(t)$ via \eqref{eq11a} and output estimation $\hat{y}(t)$ via \eqref{eq12a}\;

\textbf{End}
\end{algorithm}
\begin{algorithm}
    \caption{Implementation of HIO}
\label{algo4}
\footnotesize
\textbf{Step 1 Battery model discretization:}

In the absence of faults, the IOM-based battery system \eqref{eq0} can be expressed in a continuous linear form as:
\begin{align}
\nonumber \dot{x}(t) &= f(x(t), u(t), \omega(t)) \approx Ax(t) +  Bu(t)+\omega(t),\\
\nonumber y(t) &= h(x(t), u(t), v(t)) \approx C x(t) + Du(t) +v(t),
\end{align}
where $x_k$ denotes the battery state, $y_k$ represents the measurable output, $A$, $B$, $C$ and $D$ represent constant battery parameter matrices, modeling errors $\omega$  and measurement noises $v$ are assumed to be bounded, that is $||\omega||^2_2<\infty$
 and $||v||_2^2<\infty$\;

\textbf{Step 2 Offline observer design:}

\textit{Collect} constant battery parameter matrices $A$, $B$, $C$ and $D$\;

\begin{align}
\label{eq13}
 \dot{\hat{x}}(t) &= A\hat{x}(t) +  Bu(t) + L\left(y(t)-{\hat{y}}(t)\right),\\
\label{eq14}
 {\hat{y}}(t) &= C {\hat{x}}(t) + D u(t),
\end{align}
where the observer gain matrix $L=P^{-1}Q$ can be determined from the following linear matrix inequality \cite{YXTPE2022}:
\begin{align}
\nonumber \Pi = \begin{bmatrix}
\Pi_{1} & P & -Q^T\\
* & -\gamma^2 & 0\\
* & * & -\gamma^2
\end{bmatrix},
\end{align}
with $ \Pi_{1} = A^TP-C^TQ^T+PA-QC+\epsilon I$, $P>0$ and $Q$ denoting gain matrice to be determined, $\gamma>0$ denoting the given performance index, $\epsilon$ being a given positive scalar\;

\textbf{Step 3 Online implementation:}

\textit{Set} maximal simulation time $T_{max}$\;

\textit{Obtain} gain $L$ from \textbf{Step 2}\;

\textbf{While} $t \leq T_{max}$

\textit{Collect} terminal voltage $y(t)=U_{t}(t)$, input current $u(t) = I(t)$\;

\textit{Compute} state estimation $\hat{x}(t)$ via \eqref{eq13} and output estimation $\hat{y}(t)$ via \eqref{eq14}\;

\textbf{End}
\end{algorithm}

\begin{algorithm}
    \caption{Implementation of SMO}
\label{algo5}
\footnotesize
\textbf{Step 1 Battery model discretization:}

In the absence of faults, the IOM-based battery system \eqref{eq0} can be expressed in a continuous linear form as:
\begin{align}
\nonumber \dot{x}(t) &= f(x(t), u(t), \omega(t)) \approx Ax(t) +  B u(t) + \omega(t),\\
\nonumber y(t) &= h(x(t), u(t), v(t)) \approx C x(t) + Du(t) + v(t),
\end{align}
where $x_k$ denotes the battery state, $y_k$ represents the measurable output, $\omega$ represents the bounded modeling errors, $v$ represents the bounded measurement noises and $A$, $B$, $C$ and $D$ represent constant battery parameter matrices\;

\textbf{Step 2 Offline observer design:}

\textit{Collect} constant battery parameter matrices $A$, $B$, $C$ and $D$\;

\begin{align}
\label{eq15} \dot{\hat{x}}(t) &= A\hat{x}(t) +  Bu(t) + L\left(\tilde{e}_{y}(t)\right)+\rho sat(\tilde{e}_{y}(t)/\psi), \\
\label{eq16} \hat{y}(t) &= C \hat{x}(t) + Du(t), \\
\label{eq17} {\tilde{e}}_{y}(t) &= y(t) - \hat{y}(t),
\end{align}
 with $\psi$ representing the known sliding boundary. The $sat$ is the saturation function:
$$ sat(\varepsilon) =\left\{
\begin{aligned}
\varepsilon&, &-1\leq \varepsilon \leq 1 \\
sgn(\varepsilon)&, &\varepsilon<-1, \varepsilon>1
\end{aligned}
\right.
$$
where $sgn()$ is the sign function. With a proper given scalar $\rho$, semi-positive definite and positive definite matrices, $Q$ and $P$, the observer gain $L$ can be determined in the linear quadratic regulator (LQR) method using the Riccati equation as \cite{XCWSTVT2016}:
\begin{align}
AP+PA^T-PC^TR^{-1}CP = -Q.
\end{align}

\textbf{Step 3 Online implementation:}

\textit{Set} maximal simulation time $T_{max}$\;

\textit{Obtain} gain $L$ from \textbf{Step 2}\;

\textbf{While} $t \leq T_{max}$

\textit{Collect} terminal voltage $y(t)=U_{t}(t)$, input current $u(t) = I(t)$\;

\textit{Compute} state estimation $\hat{x}(t)$ via \eqref{eq15} and output estimation $\hat{y}(t)$ via \eqref{eq16}\;

\textbf{End}
\end{algorithm}

\begin{algorithm}
    \caption{Implementation of PIO}
\label{algo6}
\footnotesize
\textbf{Step 1 Battery model discretization:}

In the absence of faults, the IOM-based battery system \eqref{eq0} can be expressed in a continuous linear form as:
\begin{align}
\nonumber \dot{x}(t) &= f(x(t), u(t), \omega(t)) \approx Ax(t) +  Bu(t) + \omega(t),\\
\nonumber y(t) &= h(x(t), u(t), v(t)) \approx C x(t) + Du(t),
\end{align}
where $x_k$ denotes the battery state, $y_k$ represents the measurable output, $\omega$ represents modeling errors, and $A$, $B$, $C$ and $D$ represent constant battery parameter matrices\;

\textbf{Step 2 Offline observer design:}

\textit{Collect} constant battery parameter matrices $A$, $B$, $C$ and $D$\;
\begin{align}
\label{eq19}
 \dot{\hat{x}}(t) &= A\hat{x}(t) +  Bu(t) + K_{p}\left(y(t)-\hat{y}(t)\right) + \hat{\omega}(t),\\
\label{eq20}
 \dot{\hat{\omega}}(t) &= K_{i}\left(y(t)-\hat{y}(t)\right),\\
\label{eq21}
 \hat{y}(t) &= C \hat{x}(t) + D u(t),
\end{align}
where $\kappa_1$, $\kappa_2$, $\beta>0$ are given scalars, the proportional gain $K_{p}=P^{-1}\hat{P}$ and the integral gain $K_{i}=\kappa_1\kappa_2B^TP^{-1}P$ with unknown matrices $P>0$ and $\hat{P}$ can be determined from the following linear matrix inequality \cite{XGQHTVT2022}:
\begin{align}
\nonumber \begin{bmatrix}
\Xi_{11} & \Xi_{12}\\
* & -\beta I
\end{bmatrix}<0,
\end{align}
\begin{align}
&\nonumber \Xi_{11} = \begin{bmatrix}
PA+A^TP-\hat{P}C-C^T\hat{P}^T+\kappa_1\kappa_2^2(BB^TA+A^TBB^T) & P+\kappa_1\kappa_2^2BB^T-\kappa_2A^TB\\
*& -2\kappa_2B^T
\end{bmatrix},\\
&\nonumber \Xi_{12} = \begin{bmatrix}
P+\kappa_1\kappa_2^2BB^T & -\hat{P}\\
-\kappa_2 & 0\\
\end{bmatrix};
\end{align}

\textbf{Step 3 Online implementation:}

\textit{Set} maximal simulation time $T_{max}$\;

\textit{Obtain} gain $K_p, K_i$ from \textbf{Step 2}\;

\textbf{While} $t \leq T_{max}$

\textit{Collect} terminal voltage $y(t)=U_{t}(t)$, input current $u(t) = I(t)$\;

\textit{Compute} state estimation $\hat{x}(t)$ via \eqref{eq19}, \eqref{eq20} and output estimation $\hat{y}(t)$ via \eqref{eq21}\;

\textbf{End}
\end{algorithm}

\begin{table*}[!htb]
\begin{center}
\caption{A comparative analysis of various state observers}
\label{tab7}
\begin{tabular}{ l  l  l  l}
\hline
\hline
Observers & References & Features & Limitations\\
\hline
{\makecell[l]{KF\\ and its modification}} & \cite{RYCJPES2022} & Iterative gain updates  & {\makecell[l]{Laborious online \\computation effort}}\\
\hline
SVO & \cite{YXTVT2023} & Set-based estimation & Unknown uncertainties bound\\
\hline
LO & \cite{XHFSYZ2010} & Simple implementation & Uncertainties disturbance\\
\hline
HIO & \cite{YXTPE2022} & {\makecell[l]{Attenuation influence \\of modeling uncertainties}} & {\makecell[l]{Energy-bounded \\modeling uncertainties}}\\
\hline
PIO & \cite{JXJPE2016} & {\makecell[l]{Freedom augmentation \\in observer design}} & {\makecell[l]{Coupling between \\differential and \\integral compensation}}\\
\hline
SMO & \cite{SDTCS2016} & {\makecell[l]{Robust tracking \\under uncertainties}} & Chatter phenomenon\\
\hline
\hline
\end{tabular}
\end{center}
\end{table*}

\begin{figure*}[!t]
\centering
\includegraphics[width=6in]{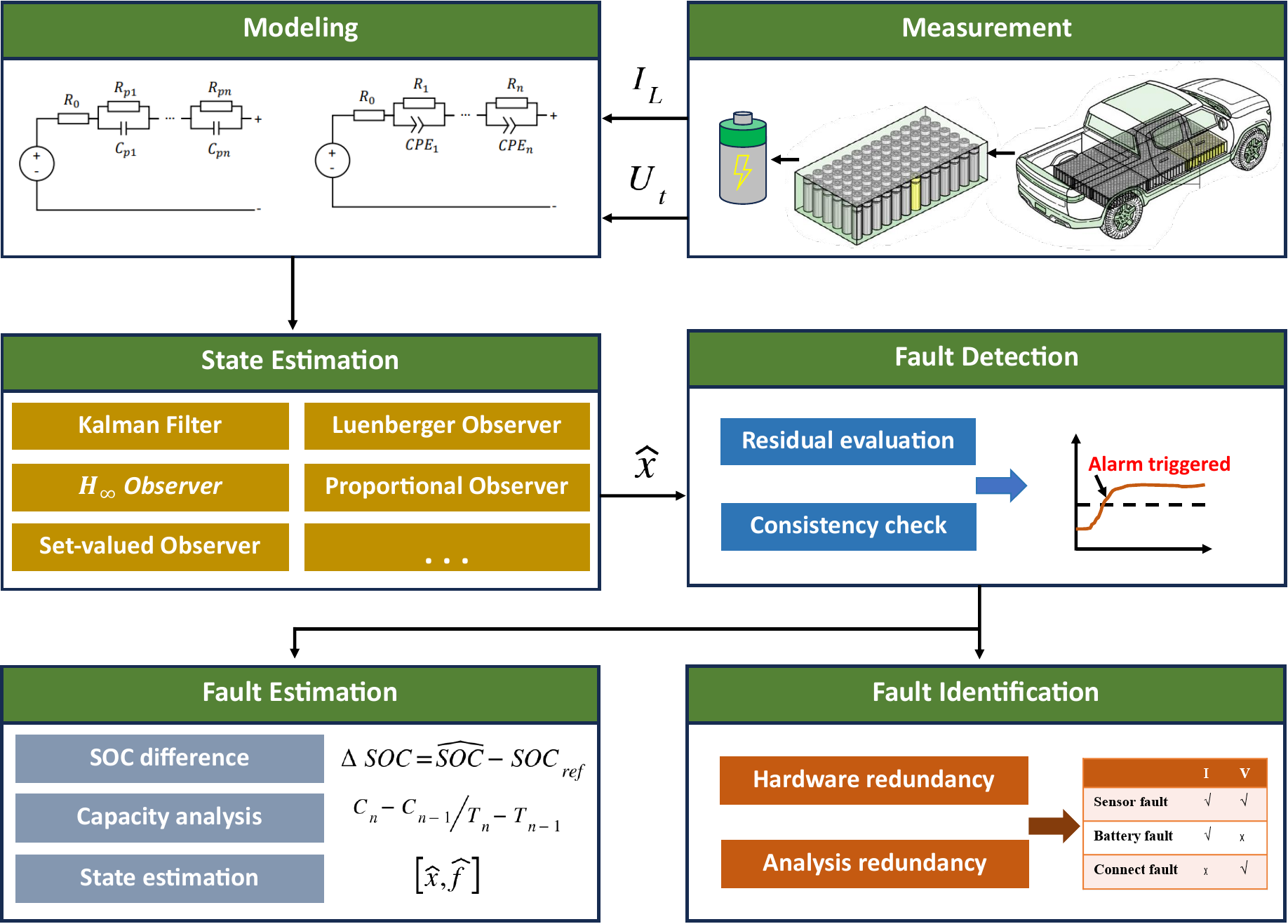}
\caption{Structure of the fault diagnosis procedure}
\label{Fig5}
\end{figure*}

\section{Fault Diagnosis}

A typical fault diagnosis procedure generally includes three parts: fault detection, fault identification and fault estimation. Fig. \ref{Fig5} illustrates the structure of the fault diagnosis procedure.

\subsection{Fault Detection}

The existing battery fault detection methods can be roughly grouped into two categories: residual evaluation for a battery cell and consistency check for a battery pack.

\subsubsection{Residual evaluation}

The basic principle for residual evaluation lies in comparing the residual between the estimation and the measurement or reference. If the generated residual deviates from the predefined threshold, a fault alarm will be triggered. For example, Hu et al. \cite{JHuIEEEJESTPE2022} applied a PSO-based method to match with measured initial system states for load current estimation. According to the residual between the estimated and measured load current, the current sensor fault can be timely detected.
Yang et al. \cite{RYCJPES2022} implemented SC fault detection via the difference between the estimated SOCs by the EKF and those computed by a Coulomb counting method.
Xu et al. \cite{YXTVT2023} developed a fault detection method through the intersection between the prediction and estimation ellipsoids obtained by set-valued observers.
The detection mechanism lies in comparing the center bias of the prediction and estimation ellipsoids with a pre-defined threshold. The presence of a sensor fault results in a discrepancy between the center of prediction ellipsoids and estimation ellipsoids, thereby activating the alarm for detection.
Kong et al. \cite{XKJES2020} estimated electrical conductivity of a separator to determine fault occurrence based on the relevant industry standard and indicate SC fault severity. Table \ref{tab8} presents a summary of the residual evaluation-based fault detection methods.

\begin{table*}[!h]
\begin{center}
\caption{A summary of residual evaluation based fault detection methods}
\label{tab8}
\begin{tabular}{l l l l}
\hline
\hline
Residual-related parameters & References &Advantages & Disadvantages\\
\hline
{\makecell[l]{Battery Measurement\\ (e.g., $U_t$, $I$)}}& \cite{JHuIEEEJESTPE2022, RXTIE2020} &  Fast reaction time& Low fault sensitivity \\
{\makecell[l]{Battery State\\(e.g., $SOC$, $U_{oc}$, $C_n$)}}  & \cite{YXTPE2022, YXTVT2023} & {\makecell[l]{Feasibility of \\quantitative assessment}} &  Slow reaction time \\
{\makecell[l]{Battery Parameter\\(e.g., $R_0$, $R_1$, $C_1$)}} & \cite{XKJES2020, KZASME2023} & High fault sensitivity & Lack of standard references\\
\hline
\hline
\end{tabular}
\end{center}
\end{table*}

The residual evaluation is commonly applied for fault detection in a battery cell. The rationale behind this is that a battery pack typically comprises numerous battery cells. Estimating the state of each cell inevitably increases computation complexity and hinders timely fault detection.

\subsubsection{Consistency check}

Consistency check involves monitoring the consistency of the estimated states, identified parameters, and measured values of the cells within the same battery pack. As a faulty battery tends to exhibit a notable deviation in measurements and estimations compared to the normal cluster, this disparity can serve as a fault indicator. For example, 
Lai et al. \cite{XLJES2020} proposed a SOC correlation-based early-stage ISC detection method for the online detection of ISCs. The correlation coefficient of SOC for normal cells exceeds the threshold, signifying high SOC consistency among these cells. Conversely, the SOC correlation coefficient for the ISC cell with a low SOC exhibits a sudden decrease, indicating low consistency between the ISC cell and the normal cells.
Ouyang et al. \cite{MOJPS2015} employed the RLS algorithm to estimate the parameters of mean-difference model. According to the characteristic parameters from $\Delta E_{parameter}$ and $\Delta R_{parameter}$, ISC detection based on the battery consistency and the statistical approach gives criteria to judge whether the cell has the risk. 
Lin et al. \cite{TLJCP2022} used the variation in the voltage difference between different cells ($d\Delta U$) as a fault index and calculated the correlation coefficients between different cell voltages and $d\Delta U$s for battery pack consistency analysis to determine fault occurrence.

\begin{figure*}[!t]
\centering
\includegraphics[width=6in]{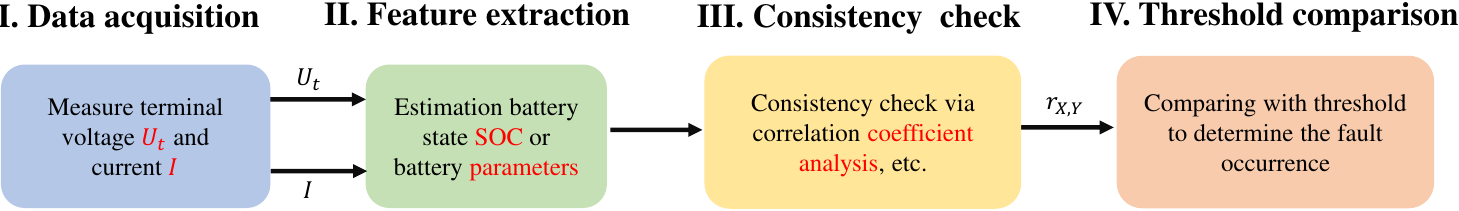}
\caption{Consistency check-based algorithm}
\label{Fig6}
\end{figure*}

Fault detection based on consistency check offers an advantage in terms of pack-level fault detection, primarily attributed to the reduction of implementation complexity. However, compared with residual evaluation, consistency check may compromise fault diagnosis sensitivity due to the interference caused by uncertainties and parameter inconsistency. The detailed consistency check-based algorithm is illustrated in Fig. \ref{Fig6}.

Regardless of residual evaluation or consistency check, fault detection can be cast into a threshold determination problem. The selection of threshold is intricately related to fault detection's false alarm rate, missing alarm rate and detection time. The use of a fixed or constant threshold may not adequately fulfil the demands of complex real-world applications. The development of an adaptive threshold that is resilient to erroneous initialization and modeling uncertainties is thus of great significance, which, to the best of our knowledge, has received limited attention in the existing battery fault diagnosis literature.

\subsection{Fault Identification}

In real EV applications, it is assumed that only one fault occurs at a time. However, one does not know the fault type before diagnosing. In order to distinguish the source and fault type, identification methods are needed. In the literature, fault identification is mainly achieved through establishing redundancy, which consists of hardware redundancy and analysis redundancy.

\subsubsection{Hardware redundancy}

\begin{figure*}[!t]
\centering
\includegraphics[width=6in]{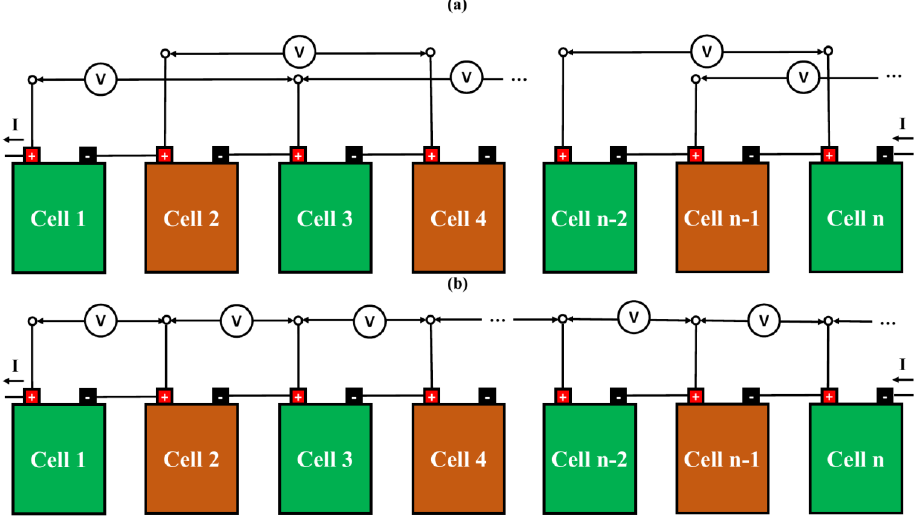}
\caption{Schematic diagram of measurement: (a) interleaved voltage measurement; (b) common voltage measurement}
\label{Fig7}
\end{figure*}

Hardware redundancy achieves fault isolation through redundant sensor implementations or sensor topology adjustments. The inclusion of redundant sensors to measure key signals, such as voltages and currents, inevitably leads to an increase in system complexity and costs. Therefore, this section primarily focuses on discussing sensor topology adjustments. For example, 
Zhang et al. \cite{KZTPE2022} devised an interleaved voltage measurement topology to distinguish voltage sensor faults from battery SC faults or open circuit faults.
Kong et al. \cite{YKJPS2019} proposed a voltage measurement that can correlate each battery and contact resistance with two different sensors for fault location and fault type identification.
Xia et al. \cite{BXJPS20162} measured the voltage summation of multiple battery cells and developed a voltage interpretation matrix to detect and isolate SC battery cell and faulty voltage sensors. The schematic diagrams for the interleaved voltage measurement and common measurement are shown in Fig. \ref{Fig7}.

\subsubsection{Analysis redundancy}

Analysis redundancy implies formulating a bank of subsystems, each associated with a specific fault, and generating residual signals by comparing the estimated system outputs with references. Each residual signal is sensitive to a particular fault and insensitive to the other faults. In this way, fault isolation can be achieved via checking the corresponding relationship between different faults and residuals by a fault isolation tree or a fault isolation table. For example, 
Liu et al. \cite{ZLCEP2016} generated voltage and temperature residuals via EKFs and employed CUSUM to evaluate the residuals for sensor fault detection and isolation.
Sidhu et al. \cite{ASTIE2015} built multiple nonlinear models representing signature faults, such as overcharge and overdischarge and generated residuals via comparing with EKF-based terminal voltage estimation. According to the conditional probability density evaluation, the fault type can be determined.
Dey et al. \cite{SDTCS2016} presented a model-based sensor diagnosis method via sliding mode observers. Fault source and bias fault can be identified based on the equivalent output error injections from observers.
Zheng et al. \cite{YZTIE2021} designed a decision tree based on the maximum SOC difference (denoted as $\text{Max}\left(\Delta SOC\right)$), the maximum change rate of $\text{Max}\left(\Delta SOC\right)$ (denoted as d$\text{Max}\left(\Delta SOC/dt\right)$) and the mean SOC for fault classification.

A comparison is performed between the hardware redundancy and analysis redundancy-based fault identification methods in terms of practicability and functionality, which is listed in Table \ref{tab9}.

\begin{table*}[!thb]
\begin{center}
\caption{A comparison between the hardware redundancy and analysis redundancy methods}
\label{tab9}
\begin{tabular}{l l l l}
\hline
\hline
 & \multicolumn{2}{c}{Hardware redundancy} & {\multirow{2}*{Analysis redundancy}}\\
& Redundant sensors& Sensor topology Adjustments&\\
\hline
Application & Cell or pack & Pack only & Cell or pack\\
Computation & Low & Low & High\\
Economic cost & High & Low & Low\\
Fault type & {\makecell[l]{Sensor fault \\or battery fault}} & {\makecell[l]{Sensor fault \\or battery fault}} &{\makecell[l]{Sensor fault \\or battery fault}}\\
System robustness & High & Easily affected by noises & High\\
Algorithm sensitivity & {\makecell[l]{Depend on \\sensor accuracy}} & {\makecell[l]{Depend on \\sensor accuracy}} & Depend on model accuracy \\
\hline
\hline
\end{tabular}
\end{center}
\end{table*}

\subsection{Fault Estimation}

Since battery voltage deviation caused by faults can sometimes be imperceptible, other deviations of battery variables such as SOC and capacity are proposed to effectively evaluate fault influence and provide a quantitative analysis of fault severity.

\subsubsection{SOC difference}

When dealing with SC fault, the reference SOC can be calculated using the Coulomb counting method since the input current is known. Due to the depletion effect of SC resistance, the SOC of a faulty battery cell will experience a reduction compared to a normal battery cell. By analyzing the SOC differences, the SC resistance can be accurately calculated. In \cite{WGTIE2019, HZTIE2023},
the SC leakage current $I_{f}$ can be calculated by the following formula:
\begin{align}
\label{eq7}
I_f= C_n \frac{\text{d}\Delta SOC}{\text{d}t}.
\end{align} Then, the SC resistance can be estimated by:
\begin{align}
\label{eq8}
R_f = \frac{U_t}{I_f}.
\end{align} To reduce the fluctuation caused by measurement noises, RLS can be selected to filter the data for higher estimation accuracy.

Similarly, in the case of a current sensor fault including bias fault, gain fault or intermittent fault, the current sensor measurement cannot accurately reflect the true input current to a battery. However, by utilizing model-based methods and incorporating feedback from correct terminal voltage measurements, the estimated state  under a faulty current sensor can still provide reliable tracking. Consequently, the reference SOC calculated based on a faulty signal will deviate from the estimated state via model-based methods, and the resultant SOC difference can then be employed for fault evaluation.

For voltage sensor faults, even though the SOC based on Coulomb counting can give an accurate state estimation, the erroneous feedback from the voltage sensor prevents the SOC from accurately tracking the true state \cite{YXTVT2023}.
Fault occurrence can be detected according to the SOC residual, however, the nonlinearity between voltage and SOC renders SOC difference-based fault estimation ineffective.

The SOC difference-based fault estimation method exhibits higher sensitivity to current-related faults than voltage-related faults. Moreover, fault estimation derived from the accumulated SOC difference may result in transient behavior and slow convergence, leading to reduced accuracy in minor battery faults.

\subsubsection{Capacity analysis}

\begin{figure*}[!ht]
\centering
\includegraphics[width=6in]{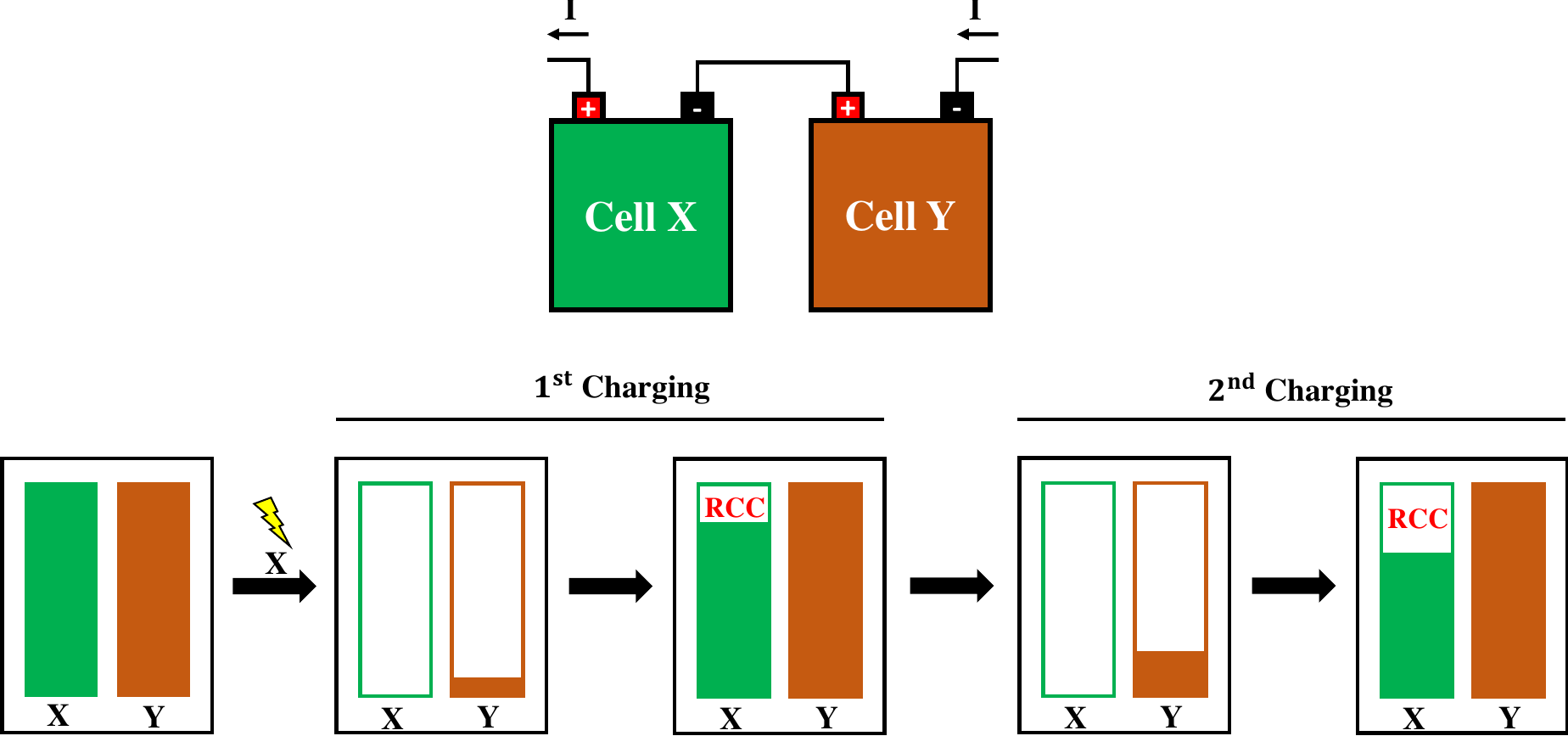}
\caption{Schematic diagram of SC estimation based on RCC}
\label{Fig8}
\end{figure*}

Capacity analysis is an effective method for fault estimation, particularly in the case of SC faults.
When an SC occurs in a battery cell, additional energy is consumed by the leakage current. This serves as a characterization of a faulty battery cell. By examining capacity-related variables such as remaining charge capacity (RCC) or incremental capacity curve, the leakage current can be calculated for the SC resistance estimation. Fig. \ref{Fig8} shows the change in RCC during the charging and discharging of the series battery cells, illustrating that SC can be well characterized by the change in RCC \cite{XKJPS2018}. Kong et al. \cite{XKJPS2018} calculated the SC-caused leakage current based on the difference between the RCCs after two adjacent charges by
\begin{align}
\label{eq9}
I_f = \frac{C_{RC, n}-C_{RC, n-1}}{T_n-T_{n-1}},
\end{align}
where $C_{RC, n}$ denotes the estimated RCC of the faulty battery cell at the end of the $n$th charge and $T_{n}$ is the time at the end of the $n$th charge.

The capacity-analysis fault estimation method is only suitable for current-related faults and its performance is significantly impacted by C-rate. To achieve a more accurate and reliable fault estimation, it is accomplished by charging or discharging a battery at a low rate within a large range which sacrifices the diagnosis time and algorithm practicability.

\subsubsection{State estimation}

State estimation-based methods to evaluate fault severity incorporate a fault index into the state vector, thereby enabling the simultaneous assessment of fault severity and other system states through an observer. For example, 
Xu et al. \cite{YXTPE2022} augmented the SC current into a battery state and introduced an $H_{\infty}$ attenuation index to suppress the influence of modeling uncertainties on the estimation.
Hu et al. \cite{JHTIE2022} proposed a multi-state-fusion ISC diagnosis method to estimate battery state including SOC and polarization voltage with load current $I$ and terminal voltage $U_t$ being set as inputs. The SC current can be estimated via the total least squares algorithm at each iteration and feedback to the faulty battery model to ensure the state estimation accuracy.
Xu et al. \cite{JXSENSOR2016} introduced a PIO to simultaneously detect and estimate the current sensor fault according to the integral of the difference between the estimated and measured terminal voltage.

State estimation-based methods for fault estimation demonstrate versatility in their applicability, as they can be employed to address various types of faults, such as SC faults, current sensor faults, and voltage sensor faults. Prior to implementing fault estimation, it becomes imperative to identify the fault type, as this step is critical for dynamic modeling of battery faults. Nevertheless, it is noteworthy that there is a scarcity of research conducted in this particular area.

\section{Challenges and Outlook}

With the rising awareness of battery safety among the general public, model-based fault diagnosis methods have been rapidly developed in the past decades. Despite great efforts made by researchers in this field, technical challenges still exist in many aspects. Some discussion and outlook are presented for future research.

\subsection{Balancing Battery Pack Modeling}
As mentioned earlier, battery packs can be categorized into series-connected packs and parallel-connected packs. Although the fault diagnosis methods reviewed can be effectively extended to series-connected battery packs, the realities of manufacturing inconsistencies necessitate the implementation of battery balancing circuits to ensure pack longevity in real-world applications. However, the significance of balancing has been barely examined in the literature. Hence, there is a need to exploit modeling for series packs that incorporates both active and passive balancing, along with voltage or SOC balancing strategies, to enhance the practical applicability of the developed fault diagnosis algorithm.

Regarding the parallel configuration, owing to the sharing terminal voltage, the parallel battery system is equivalent to having an intrinsic balancing system. Establishing an effective model for parallel-connected battery packs remains unsolved due to the coupling effect between battery cells in a pack. To tackle this issue, one can leverage the property of the same terminal voltage for diagnosis.
Moreover, acquiring branch current information is challenging or nearly impossible in parallel-connected packs. As a result, further research could be explored to investigate the feasibility of an unknown input observer in parallel packs.

\subsection{Threshold-Adaptive Fault Detection}

Conventional fault diagnosis methods rely on comparing the residual between certain variables, such as terminal voltage, SOC, etc., with a prescribed threshold to detect the occurrence of faults. The selection of the threshold value thus significantly impacts the sensitivity and robustness of fault diagnosis. A large threshold tends to increase the likelihood of missing fault alarms and to reduce SC resistance estimation accuracy. Conversely, a small threshold can increase the number of false alarms and require more iterative computation. Unfortunately, the threshold is typically selected through trial and error, leading to a lack of generalisation or repeatability. Further research is thus required to develop an adaptive threshold that maintains immunity to disturbances caused by modeling uncertainties while remaining sensitive to faults. Such an adaptive threshold would enhance fault detection performance and improve the reliability of fault diagnosis methods.

\subsection{Multi-source Fault Identification}

Apart from fault detection, some problems persist in terms of fault identification. As discussed in Section IV, LIB systems encounter various types of failures and their external characteristics at different stages exhibit minimal variations, posing difficulty in identifying and predicting the specific type of failure.
Present research on failure warning primarily concentrates on specific fault detection. In real-world applications, the fault type is often unknown in advance, yet its significance is paramount in simplifying modeling and enhancing the accuracy of fault estimation. Therefore, exploring the methods to identify fault types is a crucial subject for future research.

\subsection{Cloud-based BMS}

It is worth noting that with the increasing complexity of a BMS, its usage in EVs is encountering significant storage and computing challenges \cite{SLAE2022, KLJPS2020, DQAE2022, CXEtra2022}.
Consequently, the concept of an advanced BMS integrated with a big data cloud platform has gained considerable interest. This approach involves transferring the resource-intensive algorithms, typically requiring significant CPU and memory space in a traditional BMS, to the cloud platform to construct a cloud-based BMS. This shift helps alleviate the storage and computing pressure within EVs. Subsequently, the following three issues arise:

(1) Discrepancies in sampling rates exist compared to laboratory-based datasets: different from laboratory-based datasets, which are characterized by high sampling frequency at about 1Hz, the sampling frequency of cloud-based datasets is commonly chosen to be around 0.1Hz due to the data storage limitation. Hence, the question of whether models developed at a 1Hz sampling frequency can maintain satisfactory accuracy under reduced sampling frequency requires further investigation.

(2) Communication problems may arise when receiving and transmitting data from and to the cloud: the introduction of the cloud techniques would inevitably increase the risk of cyber-safety issue (e.g., malicious attacks, data loss), thereby the cloud-based BMS may make an incorrect evaluation on the battery states and further result in the issuance of wrong instructions. Thus, there is a desire to address the cloud-based security-guaranted battery state estimation and fault diagnosis to tackle malicious cyber attacks.

(3) Sufficient exploration of historical data for the cloud-based BMS performance enhancement: unlike the traditional BMS constrained by limited local storage, the cloud-based BMS can effortlessly analyze historical data leveraging the storage capacity of the cloud server. Exploring strategies for harnessing this data is essential for advancing our comprehension of battery evolving trends and elevating the precision of fault diagnosis.

\section{Conclusion}

This paper presents a comprehensive survey of model-based fault diagnosis methods and reviews the cutting-edge algorithms. The plant model to describe battery electrical behavior is firstly summarized from physics-based EMs and ECMs. Subsequently, to depict the nonlinear battery dynamic, a linearly parameterized state-space equation is given. Specifically, the advantages and disadvantages of various state vector configurations are discussed. Comparative studies between offline and online parameter matrix identification are performed. Moreover, the fault mechanism together with the impacts of uncertainties on state estimation is introduced. Thorough explanations are then shown for different types of observers, including their characteristics and implementations. Additionally, existing fault diagnosis methods were analyzed from three perspectives: fault detection, fault identification, and fault estimation. Finally, the challenges in model-based fault diagnosis and outlook in cloud-based BMS are summarized.

\end{document}